\documentclass[manuscript, linenumbers]{aastex63}
\received{}
\revised{}
\accepted{}
\submitjournal{AJ}

\shorttitle{IC 1590}
\shortauthors{Kim et al.}
\begin{document}

\title{The Sejong Open Cluster Survey (SOS). VII. A Photometric Study of the Young Open Cluster IC 1590}

\correspondingauthor{Beomdu Lim}
\email{blim@khu.ac.kr}

\author[0000-0001-6168-1709]{Seulgi Kim}
\affiliation{Department of Astronomy and Space Science \\
Sejong University, 209 Neungdong-ro, Gwangjin-gu, Seoul 05006, Republic of Korea}

\author[0000-0001-5797-9828]{Beomdu Lim}
%\altaffiliation{AASTeX v6+ programmer}
\affiliation{School of Space Research, Kyung Hee University, 1732, Deogyeong-daero, Giheung-gu, Yongin-si, Gyeonggi-do 17104, Republic of Korea}
\affiliation{Korea Astronomy and Space Science Institute, 776 Daedeokdae-ro, Yuseong-gu, Daejeon 34055, Republic of Korea}

%\collaboration{1}{(AAS Journals Data Scientists collaboration)}

\author[0000-0001-7801-1410]{Michael S. Bessell}
\affiliation{Research School of Astronomy \& Astrophysics, The Australian National University, Canberra, ACT 2611, Australia}
%\affiliation{AAS Journals Associate Editor-in-Chief}
%\nocollaboration{1}

\author[0000-0001-6072-9344]{Jinyoung S. Kim}
\affiliation{Steward Observatory, University of Arizona, 933 N. Cherry Ave., Tucson, AZ 85721-0065, USA}

\author{Hwankyung Sung}
\affiliation{Department of Astronomy and Space Science \\
Sejong University, 209 Neungdong-ro, Gwangjin-gu, Seoul 05006, Republic of Korea}

%\collaboration{1}{(LaTeX collaboration)}



\begin{abstract}
Young open clusters are ideal laboratories to understand star formation process.
We present deep $UBVI$ and H$\alpha$ photometry for the young open cluster IC 1590 in the center of the H II region NGC 281. Early-type members are selected from $UBV$ photometric diagrams, and low-mass pre-main sequence (PMS) members are identified by using H$\alpha$ photometry. In addition, the published X-ray source list and {\it Gaia} astrometric data are also used to isolate probable members. A total of 408 stars are selected as members. The mean reddening obtained from early-type members is $\langle E(B-V) \rangle = 0.40 \pm 0.06$ (s.d.). We confirm the abnormal extinction law for the intracluster medium.
The distance modulus to the cluster determined from the zero-age main-sequence fitting method is $12.3 \pm 0.2$ mag ($d = 2.88 \pm 0.28 ~\mathrm{kpc}$),
which is consistent with the distance $d = 2.70 ^{+0.24} _{-0.20}$ kpc from the recent {\it Gaia} parallaxes.
We also estimate the ages and masses of individual members by means of stellar evolutionary models.
The mode of the age of PMS stars is about $0.8~\mathrm{Myr}$.
The initial mass function of IC 1590 is derived.
It appears a steeper shape ($\Gamma = -1.49 \pm 0.14$) than that of the Salpeter/Kroupa initial mass function for the high mass regime ($m > 1 ~\rm{M_\odot}$).
The signature of mass segregation is detected from the difference in the slopes of the initial mass functions for the inner ($r < 2.^\prime5$) and outer region of this cluster.
We finally discuss the star formation history in NGC 281.
\end{abstract}



\section{Introduction\label{sec:intro}}

Young open clusters are the prime targets to study star formation processes.
In particular, they make it possible to obtain distance, age, luminosity and mass function because their members have the same origin.
These fundamental parameters are basic information to understand the formation and dynamical evolution of stellar clusters \citep{lada2003, sagar1986}.
Young open clusters can also be used to verify stellar evolution theory in a wide range of stellar masses \citep{sung2013} and trace the spiral structure of the Galaxy as they are the building blocks of the Galactic disk \citep{carraro1998, chen2003, piskunov2006}.
In addition, young stars in open clusters are now being targeted for probing the formation processes of exoplanets and investigating properties of protoplanetary disks where planets could be formed (e.g., \citealp{alcala2017, eisner2018, ansdell2020, sanchis2020}).

Stellar photometry is a basic tool to study the physical properties of young stellar systems \citealp{lim2015a, abdelaziz2020}.
However, since most young open clusters are distributed in the Galactic plane, it is essential to discriminate cluster members from field interlopers to determine the reliable results.
Classical T-Tauri stars (CTTSs) tend to show a prominent UV excess and H$\alpha$ emission due to accretion activity, as well as a mid-infrared (MIR) excess from their warm circumstellar disks \citep{sung1997, sung2009, megeath2012, lada1987}.
On the other hand, weak-line T-Tauri stars (WTTSs) have no or little accretion activity, but are still very bright in X-ray \citep{feigelson2003}.
Therefore, a photometric study complemented by multi-wavelength data is required to select genuine members.

The initial mass function (IMF), originally introduced by \citet{salpeter1955}, is a fundamental tool for understanding star formation processes.
If the IMF is universal in all star forming regions (SFRs), one can expect common factors to control the star formation process.
Conversely, variety in the IMFs implies that the star formation process may be dependent on the physical properties of the natal cloud.
The issue of the universality of the IMF is still an ongoing debate.
Young open clusters are ideal targets for the study of the IMF across a wide range of stellar masses because their members are less affected by stellar evolution and dynamical evolution \citep{sung1997}.
The universality of the IMF can therefore be examined through the study of many young clusters.

Star formation can be regulated by massive stars in young stellar clusters.
If massive stars (earlier than B3) are formed in a molecular cloud, HII regions can be created by their intense UV radiation.
Strong stellar winds during their lifetime and subsequent supernova events could push out the surrounding material and may create a huge void-like structure called a superbubble or a galactic chimney.
As the size of the superbubble increases, the accumulated material in the peripheral regions may induce subsequent star formation \citep{bally2008}.
%The young cluster IC 1590 is thought to be the result of such triggered star formation \citep{sato2008, megeath2002, megeath2003}.

In this context, we have dedicated a homogeneous photometric survey of young open clusters in the Galaxy \citep{lim2011, sung2013, lim2014a, lim2014b, lim2015a, lim2015b, sung2019}.
This is the seventh paper as part of this project.
The target is IC 1590 in the center of the HII region NGC 281 (Sharpless 184, PacMan Nebula, $\alpha_{2000} = 00^h52^m, \delta_{2000} = +56^\circ 4^\prime$ and $l = 123.^\circ 07, b = -6.^\circ 31$) \citep{guetter1997}.
The most prominent feature of IC 1590 is the Trapezium-like system HD 5005 composed of four O-type stars (O4V((fc)), O9.7II-III, O8.5V(n), O9.5V) \citep{sota2011}.
These massive stars are the main ionizing sources of this HII region.
The cluster is known to be young ($3.5 \sim 4.4$ Myr), and 3 kpc away from the Sun \citep{guetter1997, sharma2012, sato2008}.

The NGC 281 region has some interesting structures that are related to star formation history.
The south-western region of IC 1590 is obscured by the adjoining molecular cloud NGC 281 West.
Several X-ray sources and mid-IR sources, as well as an $\text{H}_{\text{2}} \text{O}$ maser source are detected in the region.
%\citet{sato2008} measured a trigonometric parallax of $0.355\pm0.030~ \mathrm{mas}$ for NGC 281 West using VLBI observations of the $\text{H}_{\text{2}}\text{O}$ maser source.
\citet{lee2003} studied NGC 281 West in $^{\text{12}}\text{CO(J=1-0)}$ and $^{\text{13}}\text{CO(J=1-0)}$ and found that the $^{\text{12}}\text{CO(J=1-0)}$ emission peak is very close to the $\text{H}_{\text{2}}\text{O}$ maser source.
Some indication of an outflow were also found near the maser source, implying ongoing star formation in NGC 281 West \citep{snell1990, hodapp1994, wu1996}.
An age sequence between IC 1590 and NGC 281 West was found.
Class II young stellar objects (YSOs) are located close to the cluster, while Class 0/I YSOs and molecular clumps are distributed away from the cluster \citep{megeath1997, sharma2012}.
%\citet{megeath1997} found three distinct clumps in NGC 281 West from $\text{C}^{\text{18}}\text{O(J=1-0)}$ and $\text{C}^{\text{18}}\text{O(J=2-1)}$ observations, and \citet{sharma2012} showed that Class II YSOs are closer to IC 1590 than are Class 0/I YSOs in the northern subcluster of NGC 281 West.
Furthermore, VLA $20~\rm{cm}$ continuum images reveal compact structures and edges at the north of the $^{\text{12}}\text{CO}$ peak \citep{megeath1997}.
Accordingly, NGC 281 West has been suggested as a SFR triggered by IC 1590.
Another SFR, NGC 281 East, was found from $^{\text{12}}\text{CO}$ observation by \citet{elmegreen1978}.
Three IRAS sources were embedded in this region, and a highly reddened star was found in the clump at the northern edge of NGC 281 East.
\citet{megeath1997} suggested that the star formation of NGC 281 East might have been started or strengthened by shocks.
However, it is still necessary to probe the physical associations between the SFRs and IC 1590.

%IC 1590의 IMF를 다시 연구해야할 필요성

The IMF of IC 1590 has been investigated in several studies.
\citet{guetter1997} obtained a shallow IMF ($\Gamma = -1.00 \pm 0.21$) of IC 1590 from photoelectric and CCD photometry.
From $UBVI$ photometric and H$\alpha$ spectroscopic observations by 105-cm Schmidt telescope, \citet{sharma2012} also found the slope of IMF to be flatter ($\Gamma = -1.11 \pm 0.15$) than those of the Salpeter/Kroupa IMF ($\Gamma = -1.35$; \citealp{salpeter1955, kroupa2001}). %-1.11 \pm 0.15
This shallow IMF implies that this SFR is a favorable site of massive star formation.
However, these previous studies derived the IMF using bright stars ($m>1.5 ~\rm{M_\odot}$ by \citealt{guetter1997}; $m > 2~\rm{M_\odot}$ by \citealt{sharma2012}).
Their selection of low-mass PMS stars may be less complete due to the limited data.
With our deep photometry, complemented by several archival datasets, it is possible to derive the IMF down to 1 $\rm{M_\odot}$.
In addition, the recent {\it Gaia} Early Data Release 3 (EDR3; \citealt{gaia2020}) allows us to reliably select kinematic members.
Therefore, the IMF of this cluster can be studied across a wide mass range.

The main scientific goals of this study are to derive the fundamental parameters of IC 1590 (reddening, distance, age, and the IMF) and to study the star formation process in this SFR.
To achieve these aims, we conducted deep $UBVI$ and H$\alpha$ observations and investigated the properties of IC 1590 using new optical data and additional archival data.
In Section~\ref{sec:observation}, we describe our observations, data processing, and archival datasets.
The member selection criteria are addressed in Section~\ref{sec:membership}.
In Section~\ref{sec:reddening_and_distance}, the reddening and distance of IC 1590 are determined.
The IMF is derived in Section~\ref{sec:age_and_IMF}.
Finally, we discuss the triggered star formation in this SFR and the properties of some individual stars in Section~\ref{sec:discussion} and summary in Section~\ref{sec:summary}.

\section{Observations and Archival Data\label{sec:observation}}

\begin{table*}[t!]
    \caption{Observing log}
    \label{tab:observing_log}
    \centering
    \tablewidth{500pt}
        \begin{tabular}{lccccc}
        \toprule
            Filter & Exposure time (s) & Seeing ($^{\prime\prime}$) & $k_{1\lambda}$   & $k_{2\lambda}$   & $\zeta_\lambda$ \\
        \hline
            \multicolumn{6}{l}{Maidanak AZT-22 (2005. 8. 9)} \\
            \hline
            $U$ & $30$, $600$ & $1.22$ & $0.539\pm0.022$   & $0.023$           & $21.560\pm0.011$ \\
            $B$ & $5$, $300$ & $1.09$ & $0.324\pm0.010$   & $0.026$           & $23.261\pm0.006$ \\
            $V$ & $3$, $180$ & $0.88$ & $0.231\pm0.009$   & -                 & $23.396\pm0.001$ \\
            $I$ & $3$, $60$ & $0.74$ & $0.139\pm0.010$   & -                 & $23.017\pm0.009$ \\
            \hline
            \hline
            \multicolumn{6}{l}{Kuiper $61^{\prime\prime}$ (2011. 10. 31)} \\
            \hline
            $U$ & $15$, $600$ & $1.6$ & $0.451\pm0.007$   & $0.002\pm0.003$   & $22.181\pm0.008$ \\
            $B$ & $7$, $300$ & $1.4$ & $0.253\pm0.005$   & $0.025\pm0.002$   & $23.563\pm0.007$ \\
            $V$ & $5$, $180$ & $1.4$ & $0.134\pm0.005$   & -                 & $23.566\pm0.008$ \\
            $I$ & $5$, $120$ & $1.2$ & $0.043\pm0.004$   & -                 & $22.181\pm0.008$ \\
            H$\alpha$ & $30$, $600$ & $1.2$ & $0.061$   & -                 & $19.494$ \\
        \hline
    \end{tabular}
    \tablecomments{
        $k_{1\lambda}$ is the primary atmospheric extinction coefficient, $k_{2\lambda}$ is the secondary atmospheric extinction coefficient, $\zeta_{\lambda}$ is the photometric zero-point from \citet{lim2009}.}
\end{table*}

\begin{figure}
    \centering
    \includegraphics[width=\columnwidth]{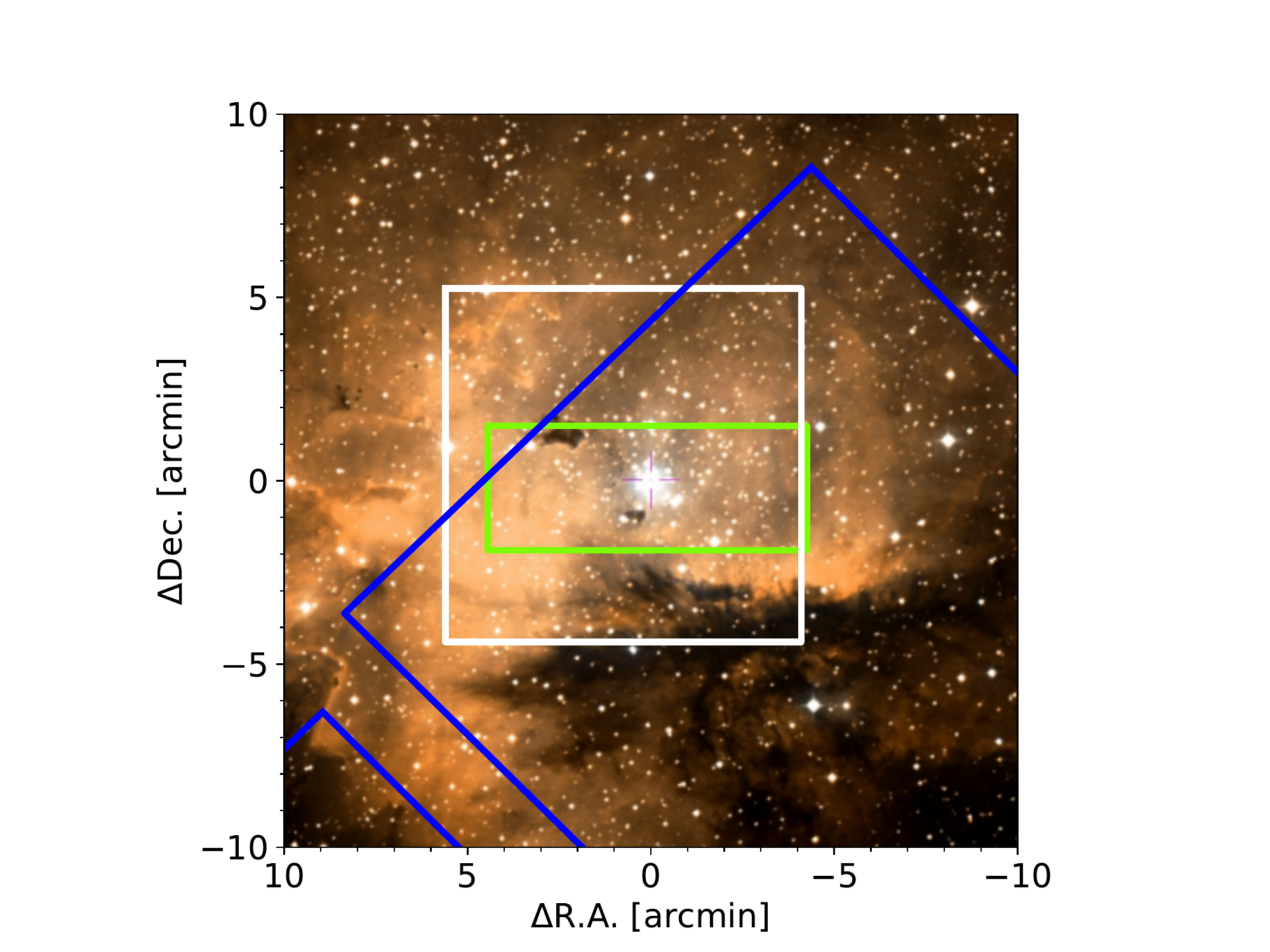}
    \caption{Optical image of NGC 281 taken from the Digital Sky Survey 2. The green rectangular and white square represent our target regions covered by the Maidanak AZT-22 observations and the Kuiper 61$^{\prime\prime}$ telescope observations, respectively. The large blue box depicts the FoV of the $Chandra$ X-ray observations. The positions of stars are relative to the coordinate R.A. = $00^{\rm{h}} 52^{\rm{m}} 49^{\rm{s}}$, decl. = $+56^\circ 37^\prime 42^{\prime\prime}$ (J2000).} % 뭔가 DSS2에 대해 footprint를 써야하나?
    \label{fig:observing_area}
\end{figure}

\begin{figure*}
    \centering
    \includegraphics[width=\textwidth]{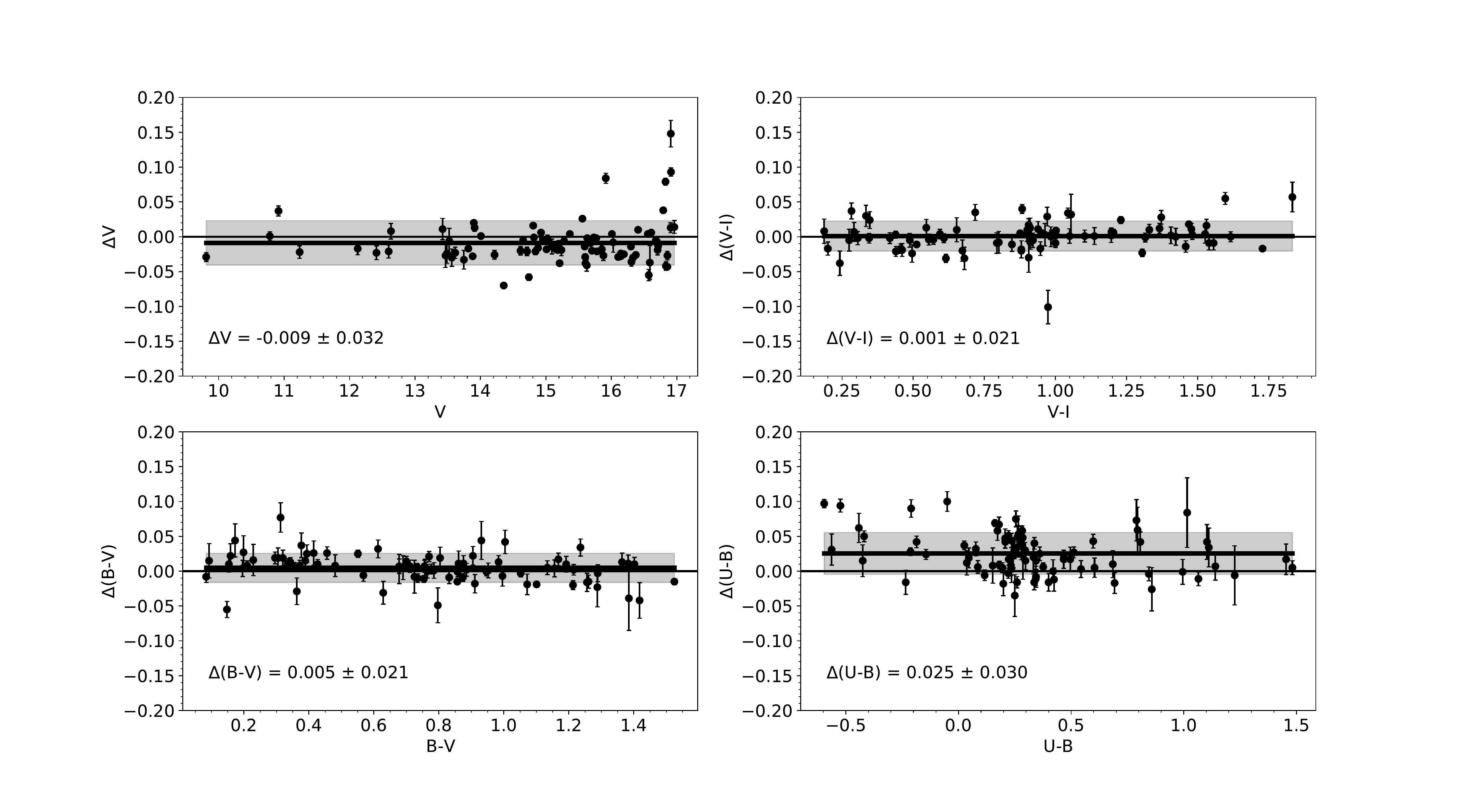}
    \caption{Comparison of two sets of photometry. The difference $\Delta$ means Kuiper $61^{\prime\prime}$ data - Maidanak data. Stars brighter than $V = 17$ mag were used in the comparison to avoid the large photometric errors of faint stars. The thick lines and shaded areas represent mean and standard deviation, respectively. The mean difference and the standard deviation between the two sets of the photometry are displayed in the lower-left corner of each panel.}
    \label{fig:compare_optical}
\end{figure*}

\subsection{Optical Photometry}
\subsubsection{\textit{Maidanak} AZT-22 $1.5 \mathrm{m}$ Observations}
The $UBVI$ CCD observations of the young open cluster IC 1590 were conducted on 2005 August 9, using the AZT-22 $1.5~\mathrm{m}$ telescope with the SITe $2000\times800$ CCD ($0.^{\prime\prime}266~\rm{pixel}^{-1}$) at the Maidanak Astronomical Observatory in Uzbekistan.
The typical seeing was $\sim 1 ^{\prime\prime}$.
We used two sets of exposure times for each band to cover both bright and faint stars.
To determine the atmospheric extinction coefficients and photometric zero-points, we also observed several SAAO standard stars \citep{menzies1991} in the Landolt standard fields and additional standard stars with extremely blue and red colors \citep{kilkenny1998}.
The transformation relation to the standard system is addressed in \citet{lim2009}.
The field of view (FoV) of the Maidanak observations is shown in Figure~\ref{fig:observing_area} (green rectangle).
We summarize our observation in Table~\ref{tab:observing_log}.

\subsubsection{\textit{Kuiper} 61$^{\prime\prime}$ Telescope of Steward Observatory}
We performed additional $UBVI$ and H$\alpha$ observations with the Kuiper $61^{\mathrm \prime\prime}$ Telescope ($f/13.5$) and Mont4k CCD of the Steward Observatory at Mt. Bigelow in Arizona on 2011 October 31.
The pixel scale is $0.^{\prime\prime}42~\mathrm{pixel}^{-1}$ in a $3\times3$ binning mode, and the FoV is about $9.^\prime7 \times 9.^\prime7$.
The exposure times and seeing for each filter are summarized in Table~\ref{tab:observing_log}.
A number of standard stars from the same catalogues used in the previous section were observed.
Transformation to the standard system is detailed in the Appendix of \citet{lim2015a}.

\begin{longrotatetable}
    \begin{deluxetable*}{lcccccccccccccccc}
        \tablecaption{Photometric Data for IC 1590\label{tab:photo_data}}
        \tablewidth{700pt}
        \tabletypesize{\scriptsize}
        \tablehead{
            \colhead{ID} & \colhead{R.A.} & \colhead{Dec.} & 
            \colhead{$I$} & \colhead{$V$} & \colhead{$V-I$} & 
            \colhead{$B-V$} & \colhead{$U-B$} & \colhead{H$\alpha$ \tablenotemark{1}} & 
            \colhead{$\epsilon_I$} & \colhead{$\epsilon_V$} & 
            \colhead{$\epsilon_{V-I}$} & \colhead{$\epsilon_{B-V}$} & 
            \colhead{$\epsilon_{U-B}$} & \colhead{$\epsilon_{\rm{H}\alpha}$} &
            \colhead{$N_{obs}$} & \colhead{H$\alpha$ emission \tablenotemark{2}}
            \\
            \colhead{} & \colhead{[$^h~^m~^s$]} & \colhead{[$^{\circ} ~^\prime ~^{\prime\prime}$]} & \colhead{[mag]} & \colhead{[mag]} & \colhead{[mag]} & \colhead{[mag]} & \colhead{[mag]} & \colhead{[mag]} & \colhead{[mag]} & \colhead{[mag]} & \colhead{[mag]} & \colhead{[mag]} & \colhead{[mag]} & \colhead{[mag]} & &
        }
        \startdata
  21 &  0 52 20.9 & +56 37 49.6 & 17.28 & 18.46 & 1.18 & 0.87 & 0.21 & -0.12 & 0.01 & 0.01 & 0.01 & 0.01 & 0.02 & 0.05 & 5 4 4 3 2 1 &   \\
  22 &  0 52 21.2 & +56 38 17.3 & 18.21 & 20.30 & 2.09 & 1.79 & - & -0.74 & 0.01 & 0.04 & 0.04 & 0.08 & - & 0.09 & 3 2 2 2 0 1 & H \\
  23 &  0 52 21.5 & +56 35 57.3 & 17.20 & 19.45 & 2.25 & 1.63 & - & -1.10 & 0.01 & 0.01 & 0.01 & 0.04 & - & 0.04 & 5 2 2 2 0 1 & H \\
  24 &  0 52 21.7 & +56 39 3.5 & 17.30 & 18.75 & 1.45 & 1.21 & 0.80 & 0.09 & 0.01 & 0.01 & 0.01 & 0.02 & 0.06 & 0.05 & 5 3 3 2 2 1 &   \\
  25 &  0 52 21.8 & +56 37 39.5 & 19.04 & 20.45 & 1.41 & 1.17 & - & - & 0.01 & 0.04 & 0.04 & 0.07 & - & - & 3 2 2 2 0 0 &   \\
  26 &  0 52 21.9 & +56 39 16.9 & 18.39 & 20.71 & 2.32 & - & - & - & 0.02 & 0.04 & 0.04 & - & - & - & 3 2 2 0 0 0 &   \\
  27 &  0 52 22.0 & +56 39 13.2 & 16.83 & 19.23 & 2.40 & 1.70 & - & -0.41 & 0.02 & 0.01 & 0.02 & 0.03 & - & 0.06 & 5 2 2 2 0 1 &   \\
  28 &  0 52 22.1 & +56 37 41.6 & 18.74 & 21.28 & 2.54 & - & - & - & 0.04 & 0.06 & 0.07 & - & - & - & 3 2 2 0 0 0 &   \\
  29 &  0 52 22.1 & +56 36 52.2 & 15.88 & 16.84 & 0.96 & 0.76 & 0.20 & 0.06 & 0.01 & 0.01 & 0.01 & 0.01 & 0.01 & 0.01 & 5 4 4 4 4 2 &   \\
  30 &  0 52 22.3 & +56 35 53.6 & 19.55 & 21.00 & 1.45 & 0.79 & - & - & 0.07 & 0.06 & 0.09 & 0.11 & - & - & 2 2 2 1 0 0 &   \\
        \enddata
        \tablenotetext{1}{H$\alpha$ means H$\alpha$ index (H$\alpha = m_{H\alpha} - {{V+I} \over {2}}$) defined by \citet{sung2000}.}
        \tablenotetext{2}{H$\alpha$ emission stars are represented as H and H$\alpha$ emission candidates are represented as h.}
        \end{deluxetable*}
\end{longrotatetable}

\subsubsection{Data Reduction and Comparison}
\begin{figure*}
    \centering
    \includegraphics[width=\textwidth]{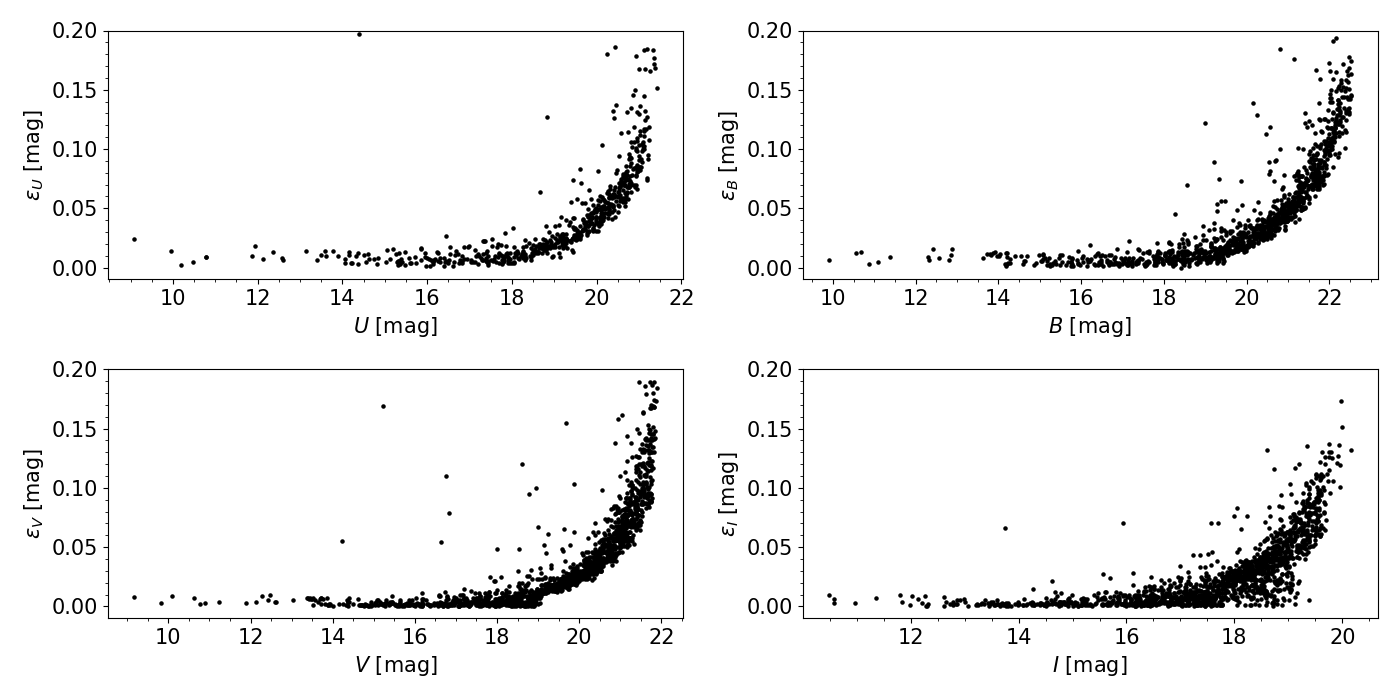}
    \caption{Phtometric uncertainties as a function of magnitude in {\it U, B, V}, and {\it I} bands for the observed region. The large errors of some stars at a given magnitude may result from the deblending of duplicated sources.}
    \label{fig:mag_err}
\end{figure*}

Pre-processing was performed using the IRAF\footnote{IRAF is developed and distributed by the National Optical Astronomy Observatories, which is operated by the Association of Universities for Research in Astronomy under a cooperative agreement with the National Science Foundation.}/CCDRED package to remove the instrumental artifacts.
The brightness of sources were measured applying a point-spread function (PSF) fitting method to the target images with IRAF/DAOPHOT \citep{stetson1987}.
A total of 731 stars and 1473 stars were detected from the Maidanak AZT-22 observations and the Kuiper 61$^{\prime\prime}$ telescope observations, respectively.
These data were transformed to the SAAO standard system as described in \citet{lim2009} and \citet{lim2015a}.
The photometric data of stars in common between these datasets were obtained from their weighted mean values, where the inverse of squared photometric errors were adopted as the weight.
We present the photometric uncertainties in Figure~\ref{fig:mag_err}.

\begin{figure*}
    \centering
    \includegraphics[width=\textwidth]{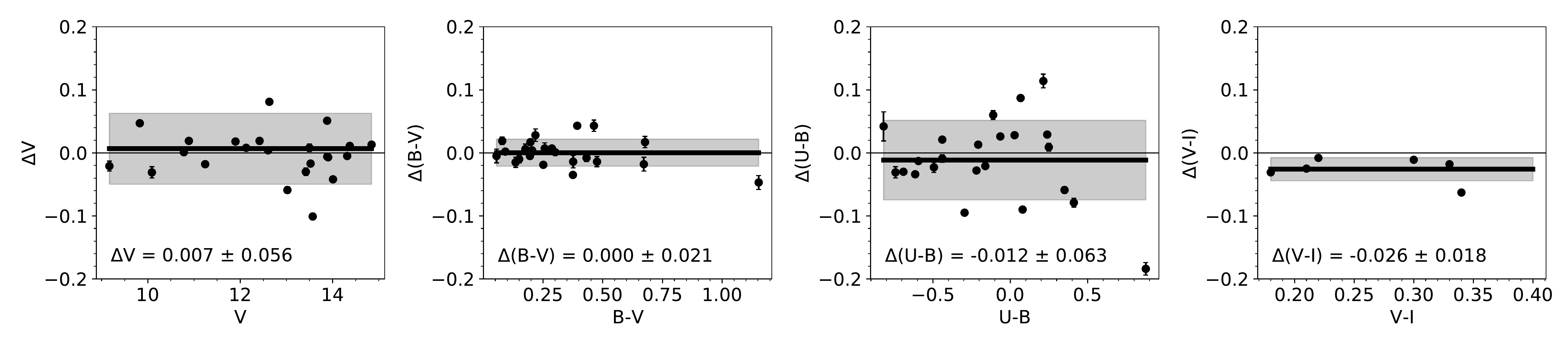}
    \caption{Comparison of our photometric data and photoelectric photometric data by \citet{guetter1997}. The thick line and shaded regions represent the mean and standard deviation, respectively. The mean difference and the standard deviation between the two sets of photometry are displayed in the lower-left corner of each panel.}
    \label{fig:compare_photoelectric}
\end{figure*}
\begin{figure}
    \centering
    \includegraphics[width=0.5\textwidth]{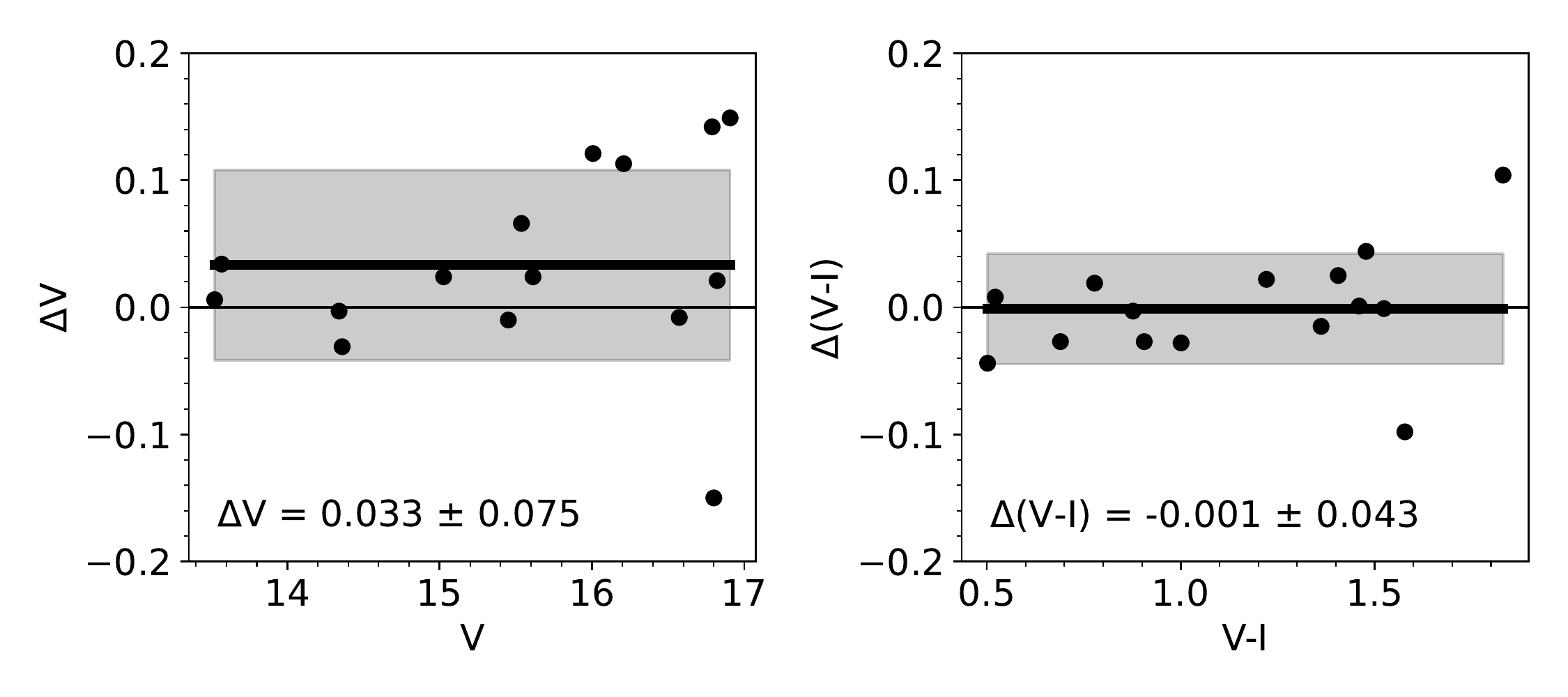}
    \caption{Comparison of our photometric data with the previous CCD data from  \citet{sharma2012}. Stars brighter than $V$ = 17 mag were used in the comparison to avoid the large photometric errors of faint stars. The thick line and shaded area represent the mean and standard deviation, respectively. The mean difference and the standard deviation between the two sets of photometry are displayed in the lower-left corner of each panel.}
    \label{fig:compare_sharma}
\end{figure}

Figure~\ref{fig:compare_optical} shows the difference between the two datasets.
These two photometric datasets show a very good consistency.
The mean difference is less than 0.01 mag for $V$, $V-I$, and $B-V$, and about 0.025 mag for $U-B$.
%\textcolor{red}{The relatively larger mean difference of 0.025 mag for $U-B$ is..}
Figure~\ref{fig:compare_photoelectric} shows the difference between our optical data and the photoelectric data of \citet{guetter1997}.
We excluded known variables and emission-line stars in the comparison.
The differences are less than 0.01 mag for $V, B-V$ and about 0.01 mag for $U-B$.
The difference for $V-I$ is about 0.03 mag, which is consistent with the expected error in the photoelectric photometry of \citet{guetter1997} which is vulnerable due to short-term variations in the sky conditions.
Figure~\ref{fig:compare_sharma} shows the difference between the CCD photometric data of \citet{sharma2012} and ours.
The mean differences are about 0.03 mag in the $V$ band and less than 0.01 mag in $V-I$.
Our data are also well consistent with the previous CCD photometric data of \citet{sharma2012}.
%Such close agreement between our data and the photoelectric photometric data supports the reliability of our data.

\begin{figure}
    \centering
    \includegraphics[width=\columnwidth]{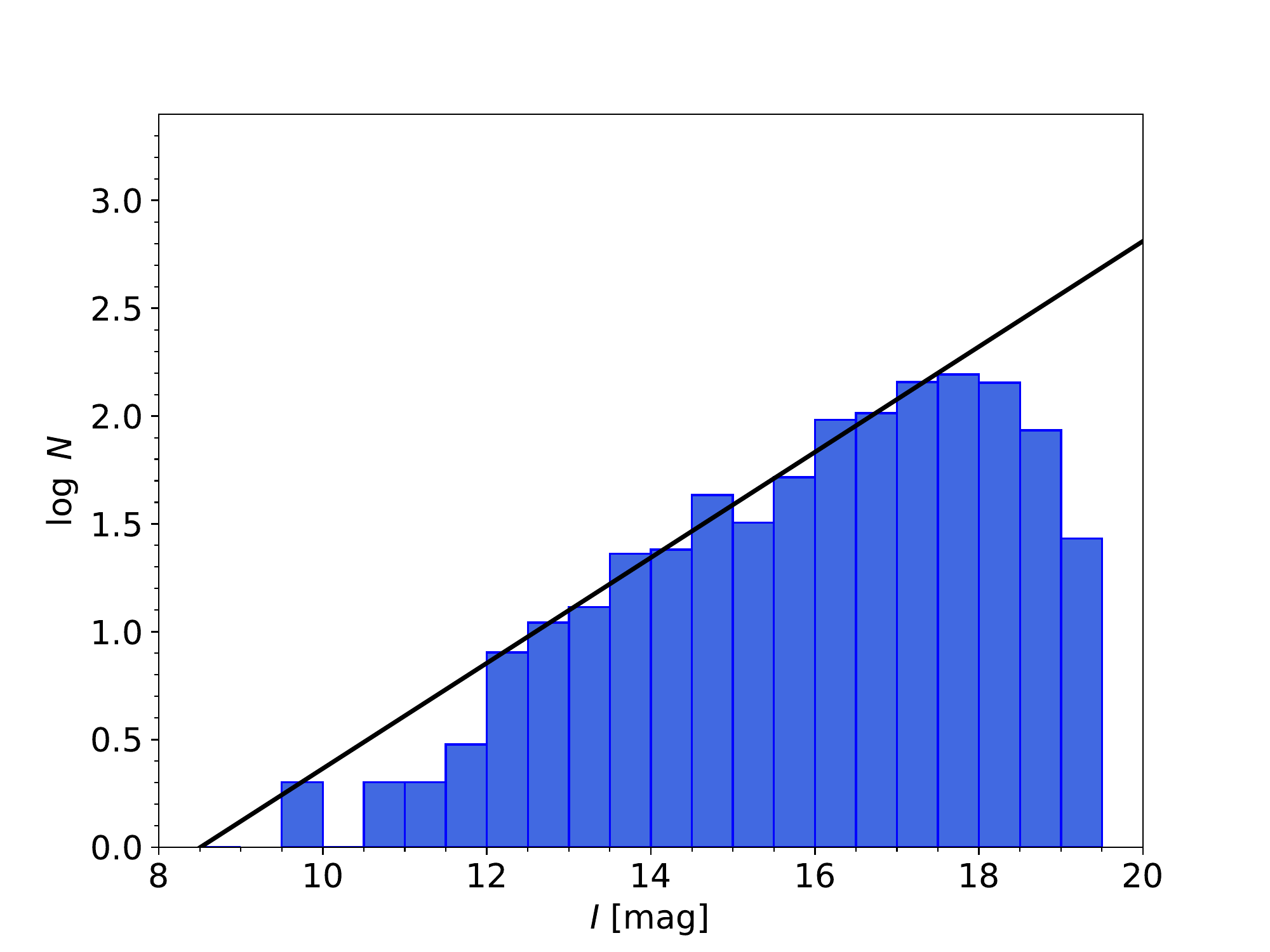}
    \caption{$I$ band luminosity function derived from all observed stars. We adopted a bin size of 0.5 mag. The turnover appears at about 17.5 mag, which is related to the completeness limit of our photometry.}
    \label{fig:luminosity_function}
\end{figure}

The brightest stars, HD 5005A and HD 5005B, were too bright to be measured reliably.
They were saturated in the Kuiper $61^{\prime\prime}$ telescope images.
Although HD 5005B was not saturated in the observation with the 2k CCD, the errors were large due to the saturated neighboring star HD 5005A.
We used instead the photoelectric photometric data of HD 5005A+B of \citet{guetter1997}.
Figure~\ref{fig:luminosity_function} shows the luminosity function of all stars in our survey region.
The least-squares method was used to determine the slope of the luminosity function between 13 and 17 mag at which our photometry is complete.
If the linear relation is extrapolated beyond 17 mag, our photometry may be 84\% complete at the turnover magnitude.

We transformed the CCD coordinates of the observed stars to the equatorial coordinates tied to the Two Micron All Sky Survey (2MASS; \citealp{cutri2003}) using several hundreds of stars in common between the two data.
A total of 833 2MASS counterparts were found in our data.
The near-infrared (NIR) data of some stars are used to investigate the extinction law.

\subsection{Archival Datasets}
\subsubsection{Gaia Data} % Gaia Collaboration 적어야하나?
Since stars in a cluster are considered to have formed from the same molecular cloud on a short timescale, cluster members share almost the same kinematic properties.
Astrometric data such as proper motions and parallaxes are useful to isolate members of the cluster from the background/foreground field stars.
The $Gaia$ EDR3 catalog \citep{gaia2020} provides accurate proper motion and parallax data for sources across the whole sky.
The nominal uncertainties reach $0.02~\mathrm{mas}$ and $0.03~\mathrm{mas}~\mathrm{yr}^{-1}$ for bright sources ($\mathrm{G} < 15~\mathrm{mag}$) and $1~\mathrm{mas}$ and $1.3~\mathrm{mas}~\mathrm{yr}^{-1}$ for $\mathrm{G} \sim 21~\mathrm{mag}$ sources.
It has been known that the {\it Gaia} EDR3 parallaxes have a global zero-point offset of $-0.017$ mas \citep{lindegren2020}.
We searched the $Gaia$ EDR3 catalog for counterparts of our observed stars, and all stars were matched in a radius of $0.^{\prime\prime}5$.
To obtain reliable results, we did not use stars with a negative parallax and compensated for the global zero-point of $0.017 ~\rm{mas}$.

\subsubsection{Chandra ACIS Data}
PMS stars are known to be bright X-ray sources ($L_X/L_{bol} \sim 10^{-3}$, \citealp{feigelson2003, linsky2007}).
Furthermore, their X-ray activity persists for a long time \citep{sung2008}.
For these reasons, PMS stars in SFRs can be well identified through X-ray imaging observations \citep{sharma2012}.
We took a list of 1185 X-ray sources in this SFR \citep{townsley2019}.
A total of 210 X-ray counterparts were found within a searching radius of $0.^{\prime\prime}7$.

\section{Member Selection\label{sec:membership}}

As the majority of open clusters are distributed in the Galactic plane, 
a large number of field interlopers are observed along the line of sight in the same FoV.
Member selection is therefore of crucial importance in the study of open clusters.

\subsection{Low-Mass Stars}

\begin{figure*}
    \centering
    \includegraphics[width=0.9\textwidth]{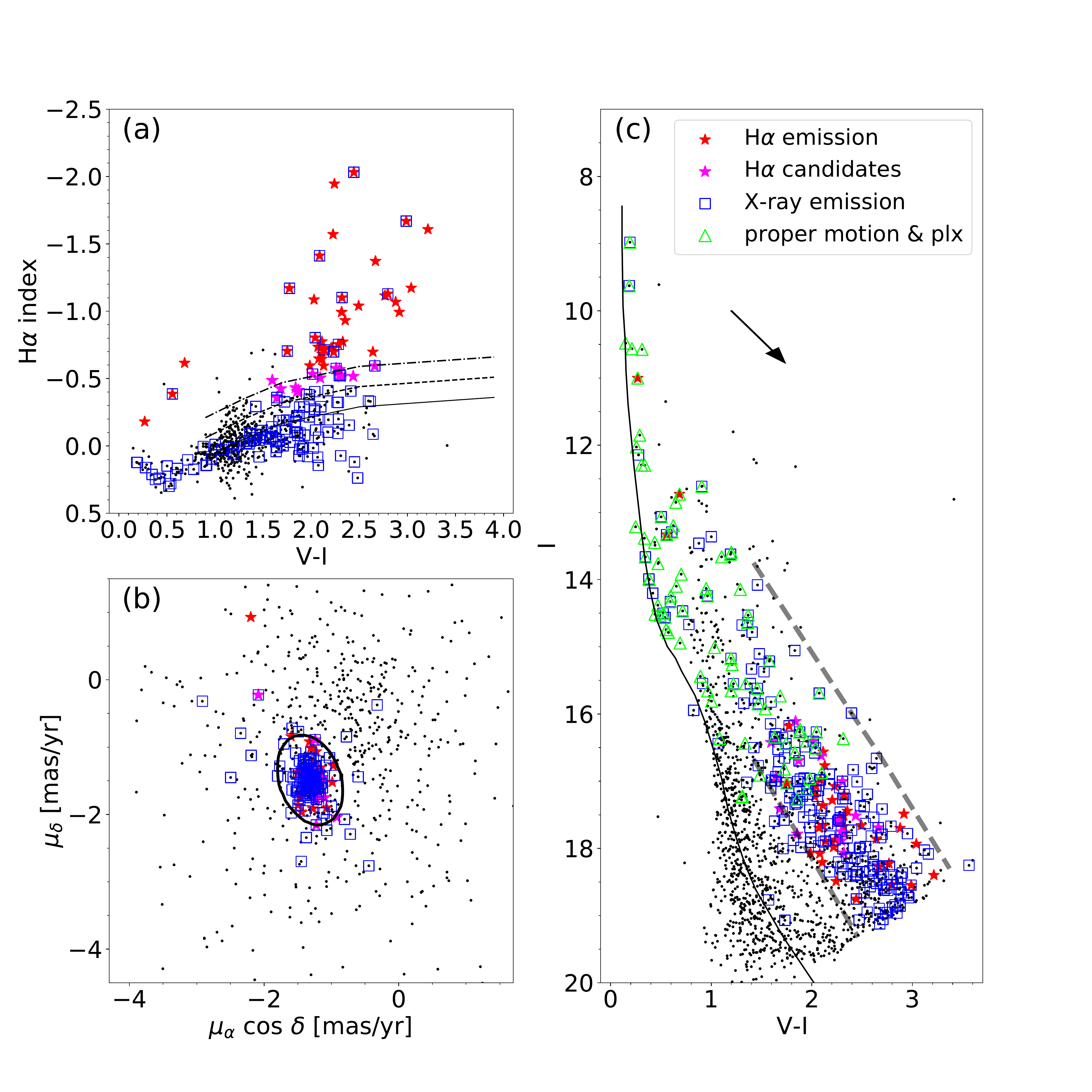}
    \caption{Member selection criteria for IC 1590. The small dots, red star symbols, magenta star symbols, blue squares, and green triangles represent all the detected stars, H$\alpha$ emission stars, H$\alpha$ emission star candidates, X-ray emission stars, and member candidates selected from the {\it Gaia} EDR3 \citep{gaia2020}, respectively. (a) H$\alpha$ index versus ($V-I$) two-color diagram. The solid fiducial line is adopted from \citet{lim2015a}. The dashed and dot-dashed lines represent the fiducial of H$\alpha$ emission candidates and H$\alpha$ emission stars. (b) Proper motion map of IC 1590. X-ray emission stars and H$\alpha$ emission stars/candidates are concentrated around ($\mu_{\alpha}\cos\delta$, $\mu_\delta$) = (-1.3, -1.5) $\rm{mas~yr^{-1}}$. (c) Color-magnitude diagram of IC 1590. The solid line represents the zero-age main-sequence relation \citep{sung2013}. We set a PMS locus between two thick dashed lines where the majority of PMS members are distributed. The reddening vector with $A_V = 1$ is shown as an arrow which is nearly parallel to the PMS locus.}
    \label{fig:membership_criteria}
\end{figure*}

\begin{figure}
    \centering
    \includegraphics[width=\columnwidth]{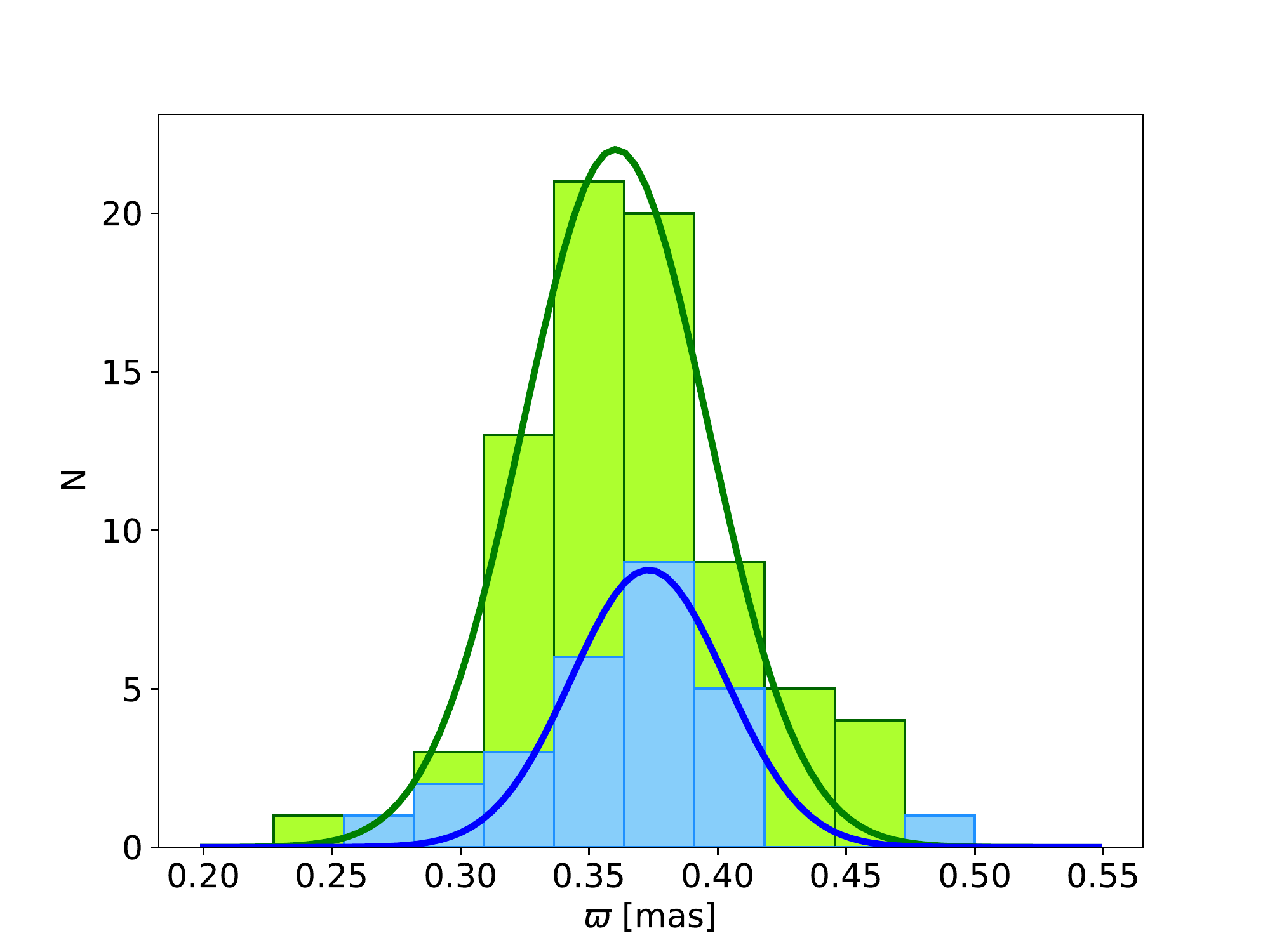}
    \caption{Distribution of parallax of X-ray emission stars (blue) and members (green) selected in Section~\ref{sec:membership}. The solid lines are fit to  Gaussian functions using the maximum likelihood method. Stars with parallaxes larger than five times the errors are used in the histogram. We corrected for the global zero-point offset of 0.017 mas \citep{lindegren2020}.}
    \label{fig:membership_plx}
\end{figure}

Material in the circumstellar disk of YSOs may accrete onto the stellar surface, manifesting in a UV excess and H$\alpha$ emission \citep{calvet1998, hartmann1999}.
Therefore, H$\alpha$ emission can be a useful indicator of CTTS in young open clusters \citep{sung1997}.
The H$\alpha$ index ($\equiv$ H$\alpha - {V+I \over 2}$) in Figure~\ref{fig:membership_criteria}(a) is a measure of H$\alpha$ emission introduced by \citet{sung2000}.
Figure~\ref{fig:membership_criteria}(a) shows the H$\alpha$ indices of the observed stars with respect to the ($V-I$) color. 
The fiducial line for H$\alpha$ emission stars is adopted from \citet{lim2015a}.
The selection criteria were set based on the distribution of $\Delta\rm{H}\alpha$,  where $\Delta\rm{H}\alpha$ means the difference between the measured H$\alpha$ indices and the fiducial line at a given $(V-I)$ color.
The $\Delta\rm{H}\alpha$ distribution appears close to a Gaussian distribution with a weak skewed wing toward negative $\Delta\rm{H}\alpha$ values.
The standard deviation of the Gaussian distribution is about 0.15 mag.
Normal stars without H$\alpha$ emission lines were found within the standard deviation from $\Delta\rm{H}\alpha = 0$, while H$\alpha$ emitting stars may be detected from the skewed wing.
Therefore, we selected $\rm{H}\alpha$ emission candidates as having $\Delta\rm{H}\alpha$ values smaller than a standard deviation ($\Delta\rm{H}\alpha < -0.15$) and $\rm{H}\alpha$ emission stars as having $\Delta\rm{H}\alpha$ values smaller than twice the standard deviation ($\Delta\rm{H}\alpha < -0.30$) from the zero value.
Some stars satisfying the criteria were excluded because of their large photometric errors.
A total of 39 H$\alpha$ emission stars and 15 H$\alpha$ emission candidates were identified.
We visually inspected these stars in the H$\alpha$ images.
In the same FoV, \citet{sharma2012} found four H$\alpha$ emission stars, which were high mass or intermediate-mass stars.
In this study, we identified a larger number of low-mass stars showing H$\alpha$ emission.

As mentioned above, identifying X-ray emission stars is the most effective way to select low-mass PMS members \citep{flaccomio2006, rauw2016, caramazza2012} because stars in the PMS stage are much brighter in X-ray than main-sequence stars \citep{prisinzano2008}.
The majority of X-ray sources as well as H$\alpha$ emission stars are found in a specific locus (hereafter PMS locus) above the field main-sequence stars at given $(V-I)$ colors (see Figure~\ref{fig:membership_criteria}(c)).
Most X-ray emission stars in the PMS locus do not show any appreciable H$\alpha$ emission, which implies that they are YSOs without an active accretion disk, at least at the time of the H$\alpha$ observations.
On the other hand, some X-ray emission stars are found below the PMS locus.
These stars may be PMS stars going through disk-bearing, or field late-type stars undergoing prolonged X-ray activity due to them being in low-mass close binary systems \citep{sung2008, fang2021}.

We also used proper motions of stars to select additional members.
Figure~\ref{fig:membership_criteria}(b) shows the proper motion of stars in the observed FoV.
We fit a 2-D Gaussian function, with the principal axis analysis, to the proper motion distribution of X-ray emission stars and H$\alpha$ emission stars that were concentrated around the mean value of $(\mu_\alpha \cos \delta, \mu_\delta) = (-1.3, -1.5) ~\rm{mas ~yr^{-1}}$
with the principal axis inclined by $19.4 ^\circ$.
The standard deviations of the major and minor axes were computed to be 0.2 and 0.1, respectively, yielding an elliptical shape that encompasses as many members as possible.
Stars with proper motions within three times the standard deviations from the mean values and associated errors smaller than $0.5 ~\rm{mas~yr^{-1}}$ were considered as member candidates (see Figure~\ref{fig:membership_criteria}(b)).
The $Gaia$ parallaxes of stars \citep{gaia2020} provide crucial constraints on cluster membership.
Figure~\ref{fig:membership_plx} displays the parallax distribution of stars.
The mean and standard deviation of X-ray emission stars are $0.36 \pm 0.03 ~\mathrm{mas}$.
Stars with parallaxes between 0.30 mas and 0.45 were selected as probable members, where we only considered parallaxes larger than three times the errors.

In addition, we considered 44 stars in the PMS locus as PMS star candidates.
They were not detected by the {\it Chandra} observation \citep{townsley2019} and do not show the H$\alpha$ emission.
Their Gaia parallaxes and proper motions are similar to those of the cluster although the associated errors are large due to their faintness.
However, we considered these stars as member candidates in this study.
The later release of the Gaia data will be able to judge their membership

Other properties of YSOs are UV excess.
The UV excess originates from hot spots in the photosphere due to the release of gravitational energy from accreting material \citep{calvet1998}.
This phenomenon that occurs in actively accreting YSOs is normally accompanied by strong H$\alpha$ emission.
Our optical data are rather limited to detect such UV excess stars because our faint YSOs have large photometric errors in the $U$ band.
However, two H$\alpha$ emission stars definitely show UV excesses in Figure~\ref{fig:BV_UB}.
In addition, some faint stars, although they have large photometric errors, are bluer in ($U-B$).

Stars satisfying more than two criteria among X-ray emission, H$\alpha$ emission, proper motion, parallax, and PMS locus are selected as the members of IC 1590.
If stars fulfill only one criterion, the stars were considered as member candidates.
We summarize the criteria of member selection as below:
\begin{enumerate}
    \item X-ray emission stars
    \item H$\alpha$ emission stars or candidates
    \item proper motion within three times the standard deviations from the mean value $(\mu_{\alpha}\cos\delta, \mu_\delta) = (-1.3, -1.5)~\mathrm{mas~yr^{-1}}$
    \item parallaxes between 0.03 mas and 0.45 mas
    \item stars in the PMS locus in HRD
\end{enumerate}
Finally, we selected 369 of the low-mass stars as members.

\subsection{Massive and Intermediate-mass Stars\label{sec:membership_OB}}
\begin{figure}
    \centering
    \includegraphics[width=\columnwidth]{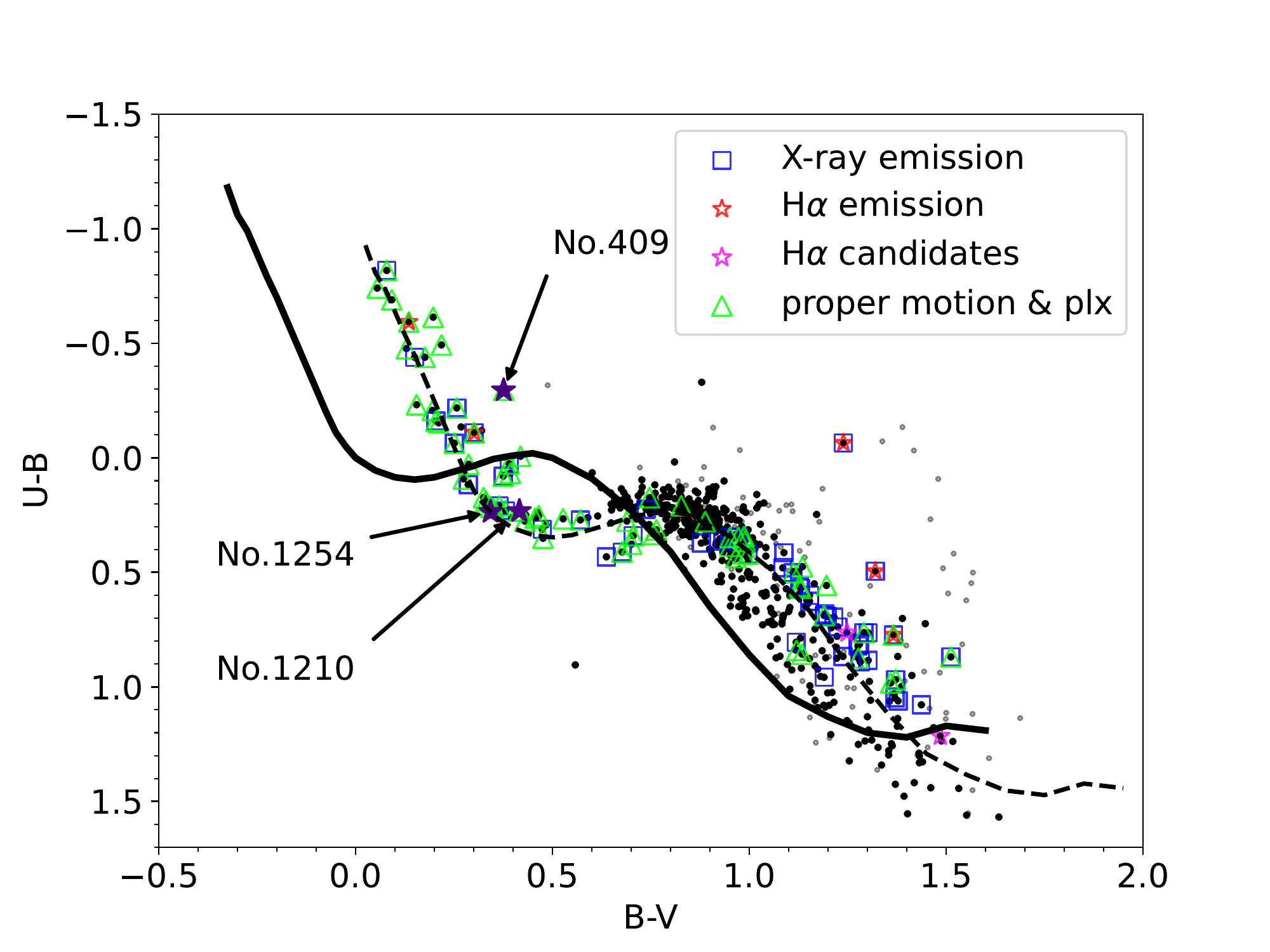}
    \caption{({\it U-B, B-V}) two-color diagram. The small open grey circles represent all detected stars and the black big dots represent stars with small errors $\epsilon_{U-B} < 0.07$ and $\epsilon_{B-V} < 0.07$ where $\epsilon_{U-B}$ and $\epsilon_{B-V}$ mean the errors of ($U-B$) and ($B-V$), respectively. The solid and dashed lines represent intrinsic color-color relation from \citet{sung2013} and the relation reddened by $E(B-V) = 0.35$, respectively. The other symbols are the same as in Figure~\ref{fig:membership_criteria}. Some notable sources among early-type stars (No.409, No.1210, and No.1254) are discussed in Section~\ref{sec:individual}.}
    \label{fig:BV_UB}
\end{figure}

\begin{figure*}
    \centering
    \includegraphics[width=\textwidth]{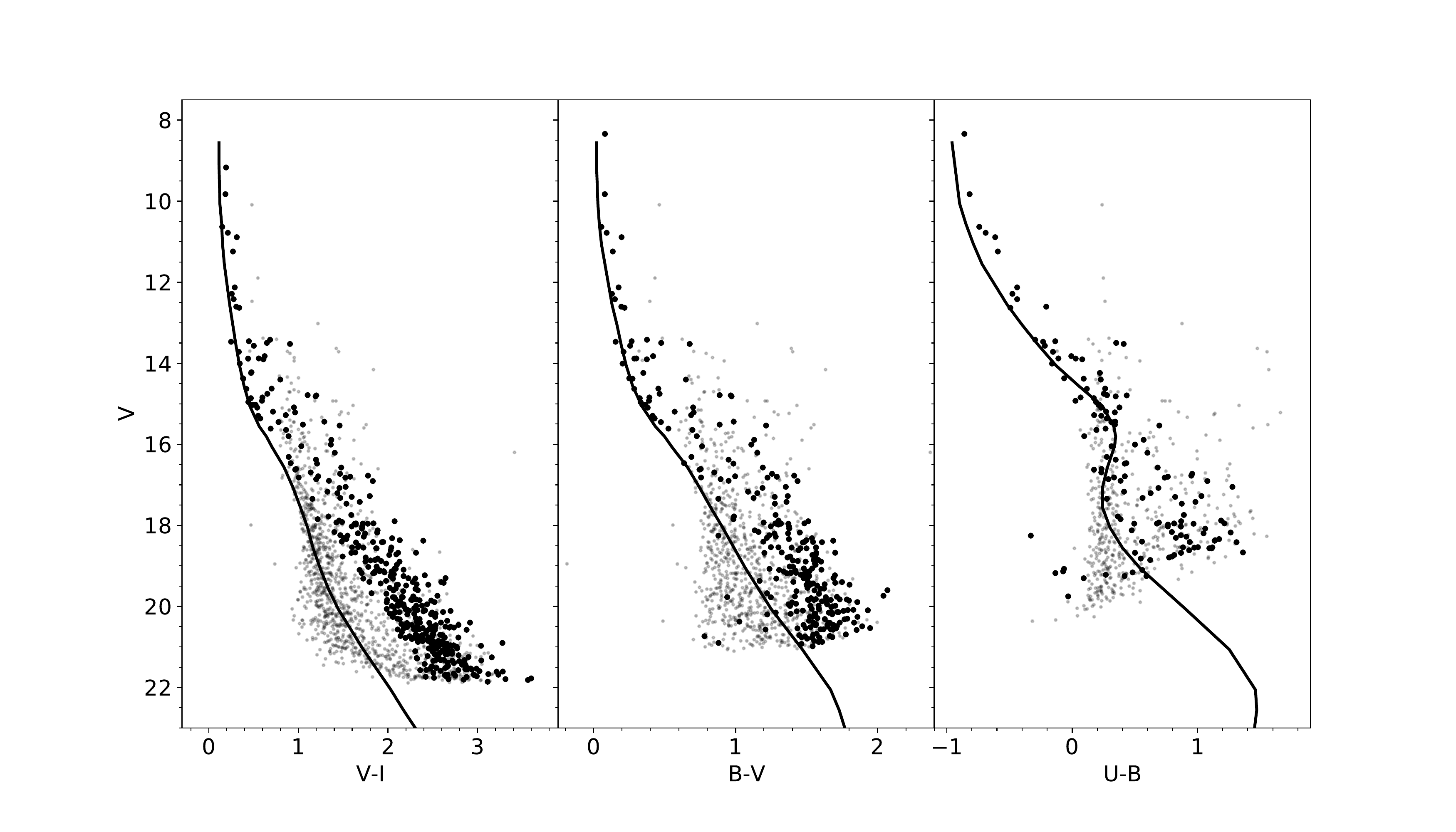}
    \caption{Color-magnitude diagrams of IC 1590. Left panel : $V-I$ vs. $V$ diagram. Middle panel : $B-V$ vs. $V$ diagram. Right panel : $U-B$ vs. $V$ diagram. The small grey dots represent all observed stars and the big black dots represent members of IC 1590. The solid lines represent the zero-age main sequence relations reddened by average reddening of members. A distance modulus of 12.3 mag was applied to the ZAMS relation (see the main text in Section~\ref{sec:distance} for detail).}
    \label{fig:CMD}
\end{figure*}

Massive O- and B-type stars are classified as probable cluster members because their lifetime is so short that they cannot migrate far away from their birth place. Such early-type stars, being luminous and blue in color, can be easily identified in photometric diagrams.
But selection of late B- to F-type members of young open clusters is very difficult because H$\alpha$ photometry is less effective to identify them (except for Herbig Ae/Be stars), and these stars are quiet in the X-ray \citep{sung2017}.
However, high precision proper motion and parallax data from the {\it Gaia} mission make it possible to identify these intermediate-mass members.
In Figure~\ref{fig:BV_UB}, stars with $B-V < 0.3$ and $V < 17$ mag corresponds to late-B-type stars, and therefore we considered these stars as early-type member candidates. Their membership was assessed by using proper motions and parallaxes.
To do this, we adopted the same criteria 3 and 4 used for the selection of low-mass members.
A total of 39 massive and intermediate-mass stars were selected as members.

Figure~\ref{fig:CMD} shows the CMD of IC 1590.
Grey and black dots represent, respectively, field stars and cluster members.
In the ($V, V-I$) CMD, PMS star population is clearly seen.
On the other hand, some PMS stars have colors bluer than those of the zero-age main sequence (ZAMS) relations in the ($V, B-V$) and ($V, U-B$) diagrams, which implies a UV excess for these PMS stars, although they have large photometric errors.
Indeed, three of these stars were identified as H$\alpha$ emission stars.
% total

\section{Reddening and Distance\label{sec:reddening_and_distance}}
\subsection{The Extinction Law\label{sec:reddening}}
The interstellar extinction can be determined in the ($U-B$, $B-V$) two-color diagram (TCD) shown in Figure~\ref{fig:BV_UB}.
The reddening vector is well established as follows \citep{sung2013};
\begin{equation}
    E(U-B) / E(B-V) = 0.72 + 0.025 \times E(B-V)
\end{equation}
We obtained the reddening of individual early-type members by comparing their observed with intrinsic colors  \citep{sung2013} along this reddening vector.

\begin{figure*}
    \centering
    \includegraphics[width=\textwidth]{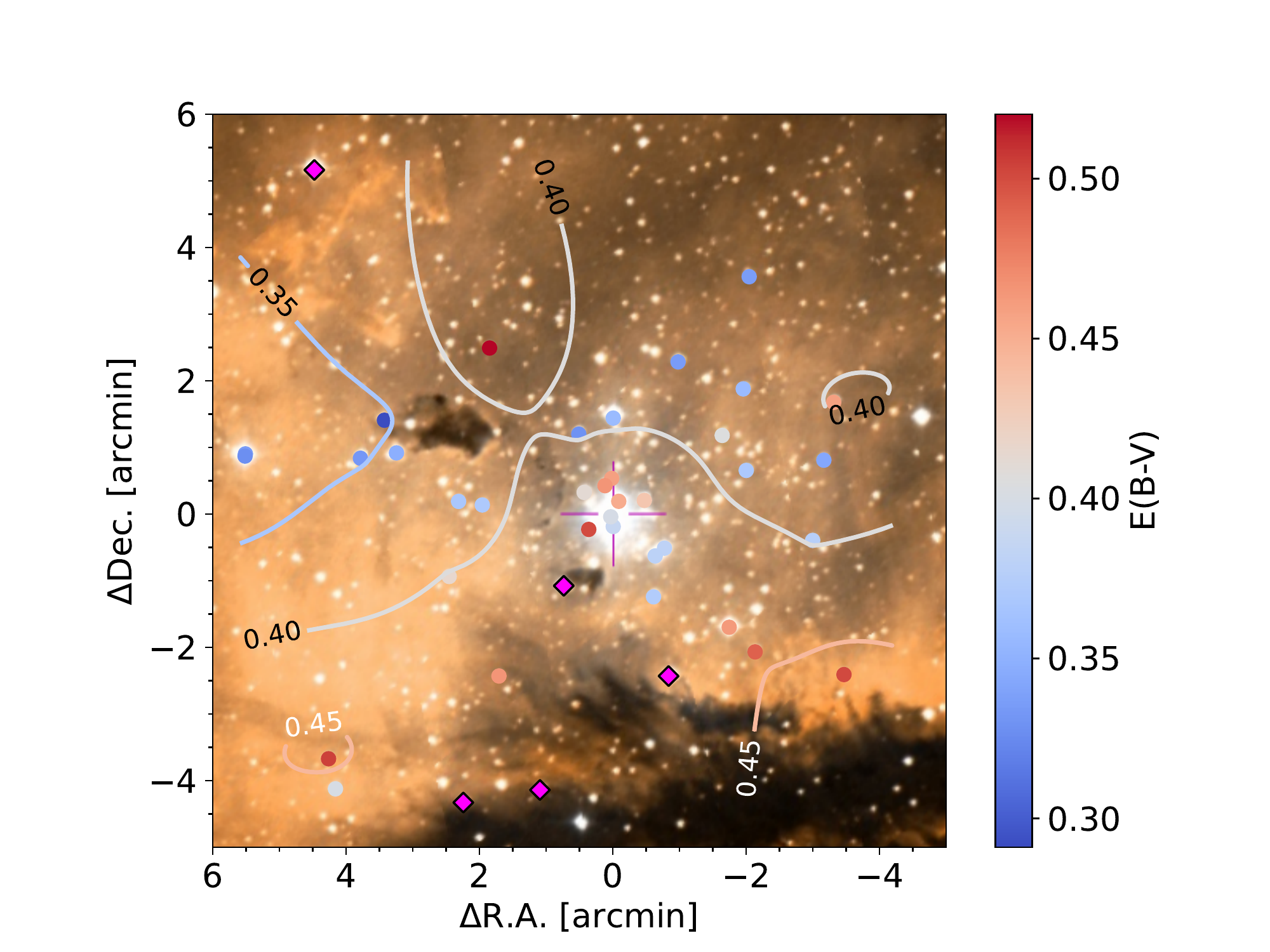}
    \caption{Interstellar extinction map. The early-type stars used for deriving the extinction map are shown as large dots. Each color of dots represents the value of $E(B-V)$. Magenta diamond symbols represent foreground stars. $E(B-V)$ of a given position in the extinction map is obtained from the weighted average of all early-type stars and the weight is adopted as $\exp{(- {r_i \over r_0})}$ where $r_i$ is the distance from the i-th early-type member and the scaling length $r_0$ is adopted as 1 arcmin.}
    \label{fig:reddening_map}
\end{figure*}

The mean reddening is $\langle E(B-V) \rangle = 0.40 \pm 0.06$ (s.d.).
There is non-negligible differential reddening given the standard deviation larger than the typical photometric errors of $0.017$ mag at $V < 20$ mag.
Figure~\ref{fig:reddening_map} shows the spatial variation of $E(B-V)$.
The southern region is more obscured by remaining material than the northern region.
The presence of the dense gas toward NGC 281 West was also confirmed from the CO observations \citep{elmegreen1978, megeath1997, lee2003}.
The reddening of individual PMS stars could not be obtained from the photometric diagrams.
Therefore, we used this map to correct the reddening of PMS stars.

Light from stars in very young open clusters is absorbed by foreground and intracluster media (see \citealp{hur2012, lim2015a}).
The properties of the line of sight extinction in given passbands are, in general, described by the total-to-selective extinction ratio ($R_V$), which depends on the size distribution of dust grains \citep{draine2003}.
The canonical value of the total-to-selective extinction ratio ($R_V=3.1$) of the diffuse ISM is generally accepted by many investigators \citep{wegner1993, lada1995, winkler1997}.
However, a somewhat lower value of $R_V=2.9$ was reported toward the Perseus arm \citep{sung2014}.
In the case of intracluster media, $R_{V,cl}$ can be high because some SFR are so young that large dust grains may still remain without having been destroyed by UV photons from massive stars.
Abnormal high extinction laws have been reported in some young open clusters (e.g., \citealt{hur2012, lim2015a, fang2021}), while others have not (e.g., \citealt{sung2017}).
The interstellar extinction due to the dust associated with IC 1590 was claimed to be larger from $R_V$ \citep{guetter1997, sharma2012}.
We corrected for the total extinction $A_V$ of each star as follows;
\begin{eqnarray}
    A_V & = & A_{V,fg} + A_{V,cl} \\ 
    & = & E(B-V)_{fg} \times R_{V,fg} + E(B-V)_{cl} \times R_{V,cl}.
    \label{eq:extinction}
\end{eqnarray}
where $A_{V, fg}$, $E(B-V)_{fg}$, $R_{V,fg}$, $A_{V,cl}$,  $E(B-V)_{cl}$, and $R_{V,cl}$ represent the foreground extinction, foreground reddening, foreground $R_V$, intracluster extinction, intracluster reddening, and intracluster $R_V$, respectively.
%Some young open clusters show an abnormal extinction law (e.g., \citealt{hur2012, lim2015a}), while others do not (e.g., \citealt{sung2017}).
%The interstellar extinction due to the dust associated with IC 1590 was claimed to be larger from $R_V$ \citep{guetter1997, sharma2012}.

\begin{figure*}
    \centering
    \includegraphics[width=\textwidth]{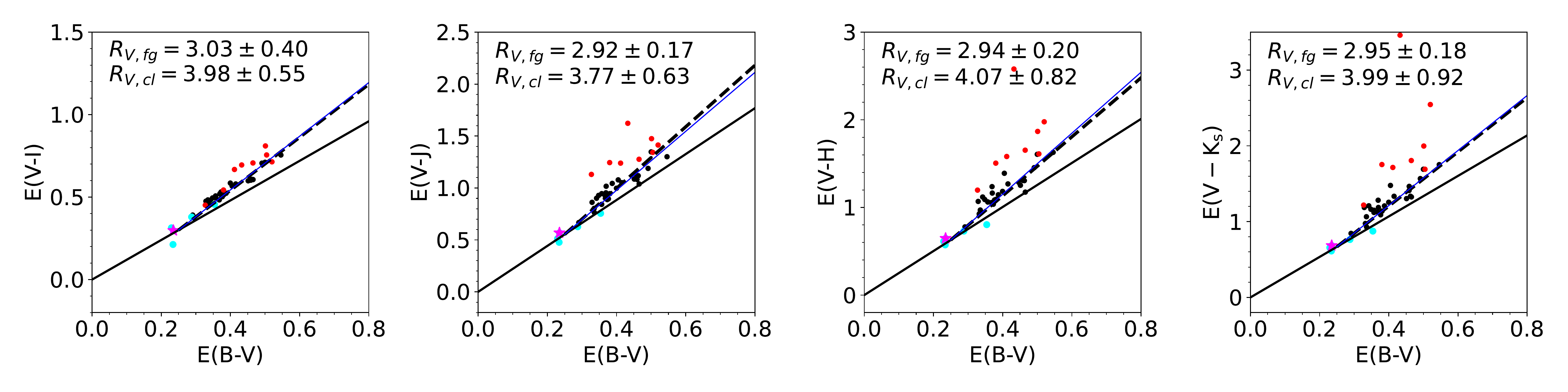}
    \caption{Color excess ratio diagrams. Black, cyan, and red colors represent, respectively, the early-type stars used in the reddening estimation, foreground stars, and NIR excess stars. The black solid and dashed lines represent the adopted color excess ratios for $R_{V,fg} = 2.94$ and $R_{V,cl} = 3.94$, respectively. The blue thin solid line shows the relation between $E(B-V)$ and $E(V-\lambda)$ for each band.}
    \label{fig:reddening}
\end{figure*}

To check whether the extinction law of IC 1590 is abnormal or not, the properties of extinction were investigated using color excess ratios.
%To determine the extinction law using the relation between $R_V$ and color-excess ratios \citep{guetter1989}, 
We estimated the intrinsic colors of early-type stars in the NIR wavelength passbands interpolating their reddening-corrected $B-V$ to the intrinsic color relations between ($B-V$) and ($V-\lambda$), where $\lambda$ is the $I$, $J$, $H$, or $K_s$ magnitude \citep{sung2013}.
The color excesses $E(V-\lambda)$ were calculated by comparing the observed colors with the intrinsic colors.
Figure~\ref{fig:reddening} shows the color excess ratios of various passbands with respect to $E(B-V)$.
$R_{V,cl}$ was obtained from the color excess ratio of $E(V-\lambda)$ to $E(B-V)$.
In some regions, $R_V$ from the $I$ band showed a considerably different value than those from the NIR $JHK_s$ bands (e.g., \citealt{hur2015}), so we did not use the $I$ band for obtaining $R_V$.
There are several stars with an IR excess emission (see Figure~\ref{fig:reddening}).
Their IR excess may originate from circumstellar material.
Hence, we cautiously excluded stars with $E(V-\lambda)$ larger than 2.5 times the standard deviation from the rest.
The mean $R_{V, cl}$ obtained from the NIR color excess is $3.94 \pm 0.34$ (s.d.).
%Our result well supports the abnormal reddening toward this cluster reported by previous studies \citep{guetter1997, sharma2012}.}

In our FoV, there are five foreground early-type stars.
Their mean reddening is $E(B-V)_{fg} = 0.24 \pm 0.02$ (s.d.).
From these stars, we can examine the foreground extinction law in the same way.
The $R_{V,fg}$ is $2.94 \pm 0.10$ (s.d.), which is normal value in the direction of the Perseus Arm \citep{sung2014}.


\subsection{Distance\label{sec:distance}}

\begin{figure*}
    \centering
    \includegraphics[width=\textwidth]{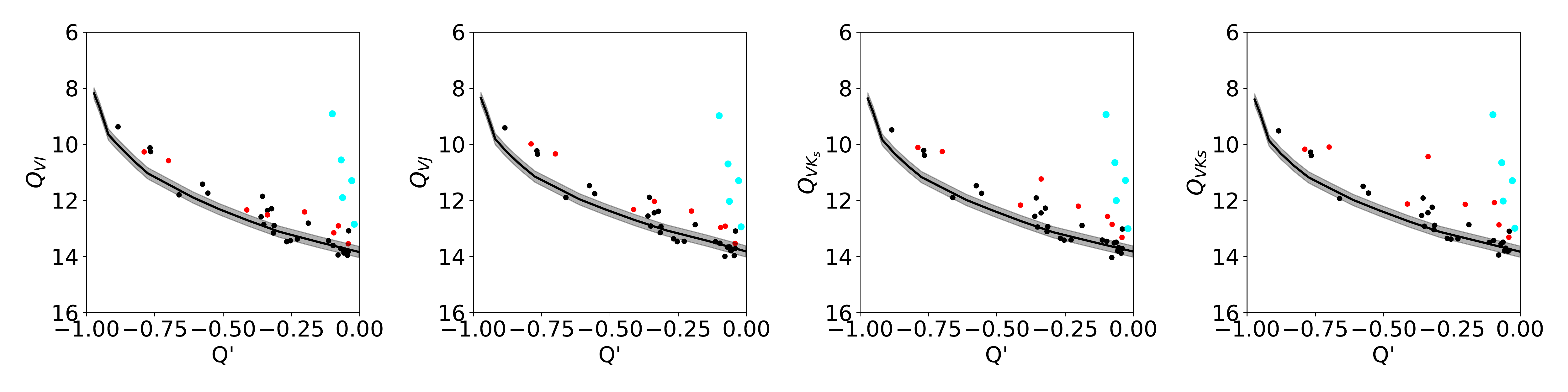}
    \caption{ZAMS fitting to the early-type main-sequence stars of IC 1590. The ZAMS relations of \citet{sung2013} were fit to the lower ridge line of the members. The solid line and shaded region represent the ZAMS relation with the distance modulus of $12.3 \pm 0.2$ mag. The color of symbols are the same as in Figure~\ref{fig:reddening}.}
    \label{fig:distance}
\end{figure*}

%The $Gaia$ astrometric data provide us with very precise proper motion as well as parallax data.
The colors and magnitudes of young main-sequence stars are empirically well established \citep{sung2013}.
We determined the distance of IC 1590 by comparing the CMD of the early-type main sequence members with the ZAMS relations of \citet{sung2013}.
\citet{johnson1953} introduced a reddening-free {\it Q} value, $Q = (U-B) - R_V \times E(B-V)$, and investigated the {\it Q} value for O- and B-type stars (e.g., \citealp{johnson1953, vandenbergh1968}).
To minimize the effect of reddening, we used reddening-free quantities and a modified Johnson $Q$ in the distance determination.
These quantities are defined as follows \citep{sung2013}.
\begin{eqnarray}
        Q_{VI} &=& V - 2.45 \times (V-I) \\
        Q_{VJ} &=& V - 1.33 \times (V-J) \\
        Q_{VH} &=& V - 1.17 \times (V-H) \\
        Q_{VK_s} &=& V - 1.10 \times (V-K_s) \\
        Q^\prime &=& (U-B) - 0.72 \times (B-V) - 0.025 \times  E(B-V)^2
\end{eqnarray}
Figure~\ref{fig:distance} shows the reddening-free CMDs of early-type stars.
Because a large fraction of high mass stars constitute binary systems \citep{sana2012} as well as the effect of rapid evolution, high mass stars may be brighter than the ZAMS at a given effective temperature \citep{lim2015a}.
Therefore, we fit the ZAMS to the lower ridgeline of the mid- to late-B type main-sequence stars within 0.2 mag.
The estimated distance modulus is $12.3 \pm 0.2 ~\mathrm{mag}$, which is equivalent to the distance of $d = 2.88 \pm 0.28 ~\mathrm{kpc}$.

We also determined a distance of IC 1590 from the inversion of the parallaxes of members.
Stars with a parallax larger than five times the error were used to determine the distance.
We fit a Gaussian distribution to the parallax distribution of members (Figure~\ref{fig:membership_plx}) and obtained $\varpi = 0.37 \pm 0.03$ mas (s.d.).
As a result, the distance is computed to be $2.70^{+0.24}_{-0.20}$ kpc.
The distance from the parallaxes is well consistent with that from the ZAMS fitting.
Our result is also in a good agreement with the distance $d = 2.82\pm0.24 ~\mathrm{kpc}$ from radio VLBA astrometry of an $\mathrm{H_2O}$ maser in NGC 281 West \citep{sato2008}.

\section{Age and IMF\label{sec:age_and_IMF}}
\subsection{Hertzsprung-Russell Diagram}
To construct the Hertzsprung-Russell Diagram (HRD), we corrected the reddening and total extinction in visual band, according to Equation~\ref{eq:extinction}.
While the reddening values of individual early-type members were applied to their observed colors, the values inferred from the reddening map were adopted for the reddening correction of lower-mass stars.
We also applied the distance modulus of 12.3 mag to the extinction-corrected visual magnitude.
The effective temperature ($T_{eff}$) of O-type stars was inferred from the spectral type-$T_{eff}$ relation, while that of the other stars was estimated by interpolating their intrinsic colors to the color-$T_{eff}$ relations.
The bolometric correction values were obtained from the BC-$T_{eff}$ relation.
These relations are obtained from \citet{sung2013}.

\begin{figure}
    \centering
    \includegraphics[width=0.7\columnwidth]{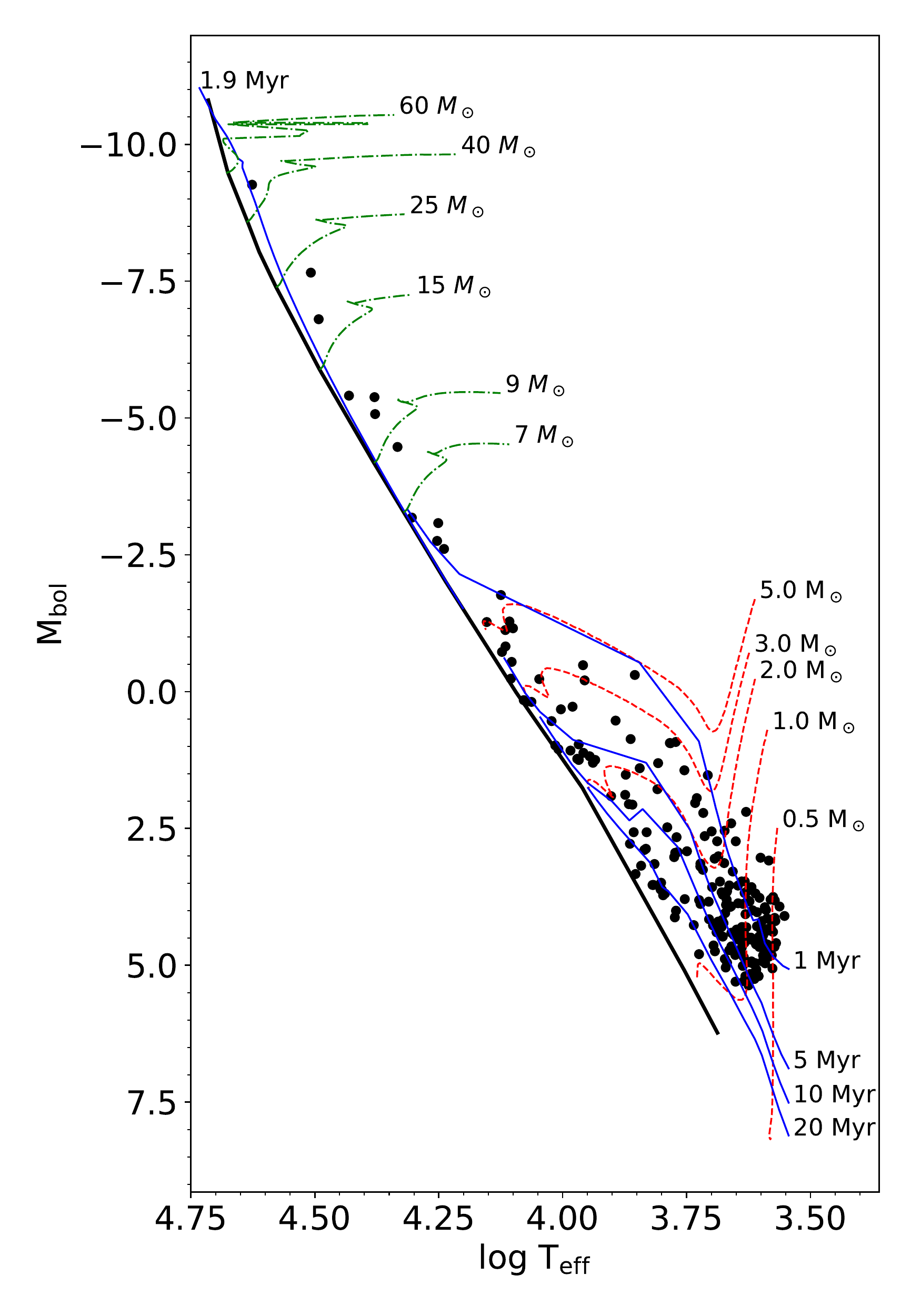}
    \caption{Hertzsprung-Russell diagram of IC 1590. The solid line in the upper part of the HRD is the ZAMS from \citet{ekstrom2012}. Some evolutionary tracks for post-main sequence \citep{ekstrom2012} and PMS stars \citep{siess2000} are plotted by green dot-dashed lines and red dahsed lines, respectively. Isochrones for four different ages interpolated from these two evolutionary models are also superimposed on the HRD (blue solid lines).}
    \label{fig:HRD}
\end{figure}

We present the HRD of IC 1590 in Figure~\ref{fig:HRD} with several stellar evolutionary tracks \citep{ekstrom2012} and PMS evolution models \citep{siess2000}.
Stars with masses larger than $7~\rm{M_\odot}$ are in the main-sequence stage.
The age of this cluster inferred from the main-sequence turn-off is about $1.9 ~\rm{Myr}$.
However, the stars in the HD 5005AB system were not spatially resolved, its luminosity may therefore be overestimated.
Hence, the age of these stars may be younger than $1.9~\rm{Myr}$.
Most of the PMS members are evolving along Hayashi tracks or in the Kelvin-Helmholz contraction phase.
The upper mass range of these PMS members appears to be $5~\rm{M_\odot}$ and the majority of PMS stars are younger than $5~\rm{Myr}$.

%\subsection{Mass and Age}

\begin{figure}
    \centering
    \includegraphics[width=\columnwidth]{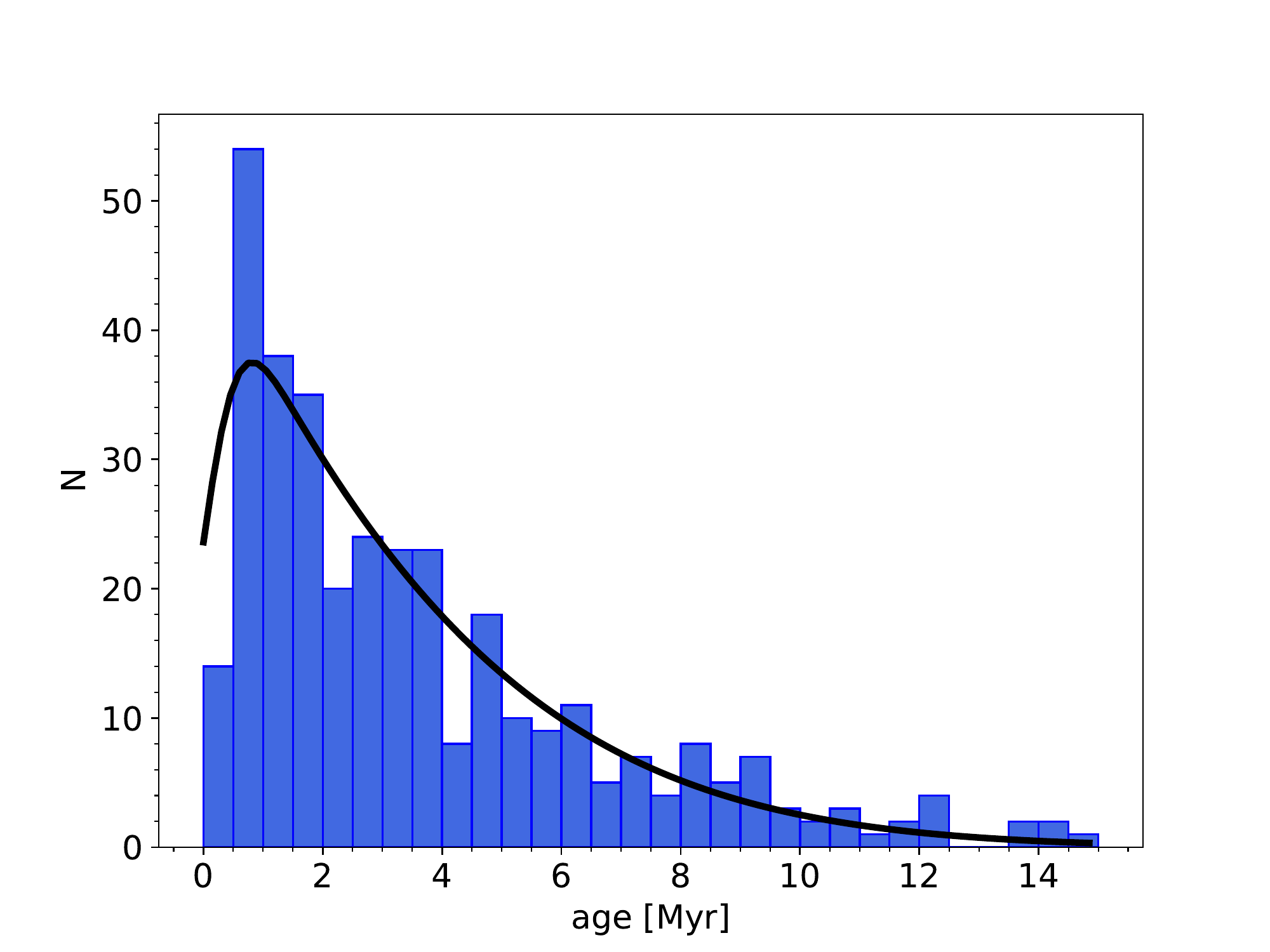}
    \caption{Age distribution of PMS stars with $m < 2 ~\rm{M_\odot}$. Black solid line represents the fit skewed Gaussian function. The mode age is $0.8$ Myr.}
    \label{fig:age_distribution}
\end{figure}

The masses and ages of individual members were derived by interpolating the $T_{eff}$ and $M_{bol}$ of each star to those from the evolutionary tracks for the main-sequence \citep{ekstrom2012} and PMS \citep{siess2000}.
We present the age distribution of PMS stars with $m < 2 ~\rm{M_\odot}$ in Figure~\ref{fig:age_distribution}.
The age distribution shows a large spread because the ages of PMS stars are almost certainly overestimated due to imperfect correction for differential reddening as well as internal extinction of disk-bearing stars, variability, and photometric errors for faint stars \citep{sung1997, sung2004}.
We fit a skewed Gaussian function to the age distribution.
The representative age of this cluster was estimated to be $0.8~\rm{Myr}$ from the mode value of the age distribution.
The ages of PMS stars ranges from 0.7 Myr to 8.4 Myr at 10\% and 90\% of the cumulative distribution.

\subsection{Initial Mass Function\label{sec:initial_mass_function}}
The mass function can be derived from the distribution of masses of individual stars.
The IMF is expressed as the number of stars in a given logarithmic mass bin in a projected area $S$, i.e. $\xi \equiv {N \over \Delta \mathrm{log} m \cdot S}$, where $N$ and $\Delta \mathrm{log}~ m$ represent the number of stars within a given mass bin and the size of the logarithmic mass bin, respectively.
The IMF is approximated as a linear relation of stellar mass in a logarithmic scale, i.e. $\log \xi \propto \Gamma \times \log m$.
\citet{salpeter1955} obtained the slope $\Gamma = -1.35$ for the stars in the Solar neighbourhood.

\begin{figure}
    \centering
    \includegraphics[width=\columnwidth]{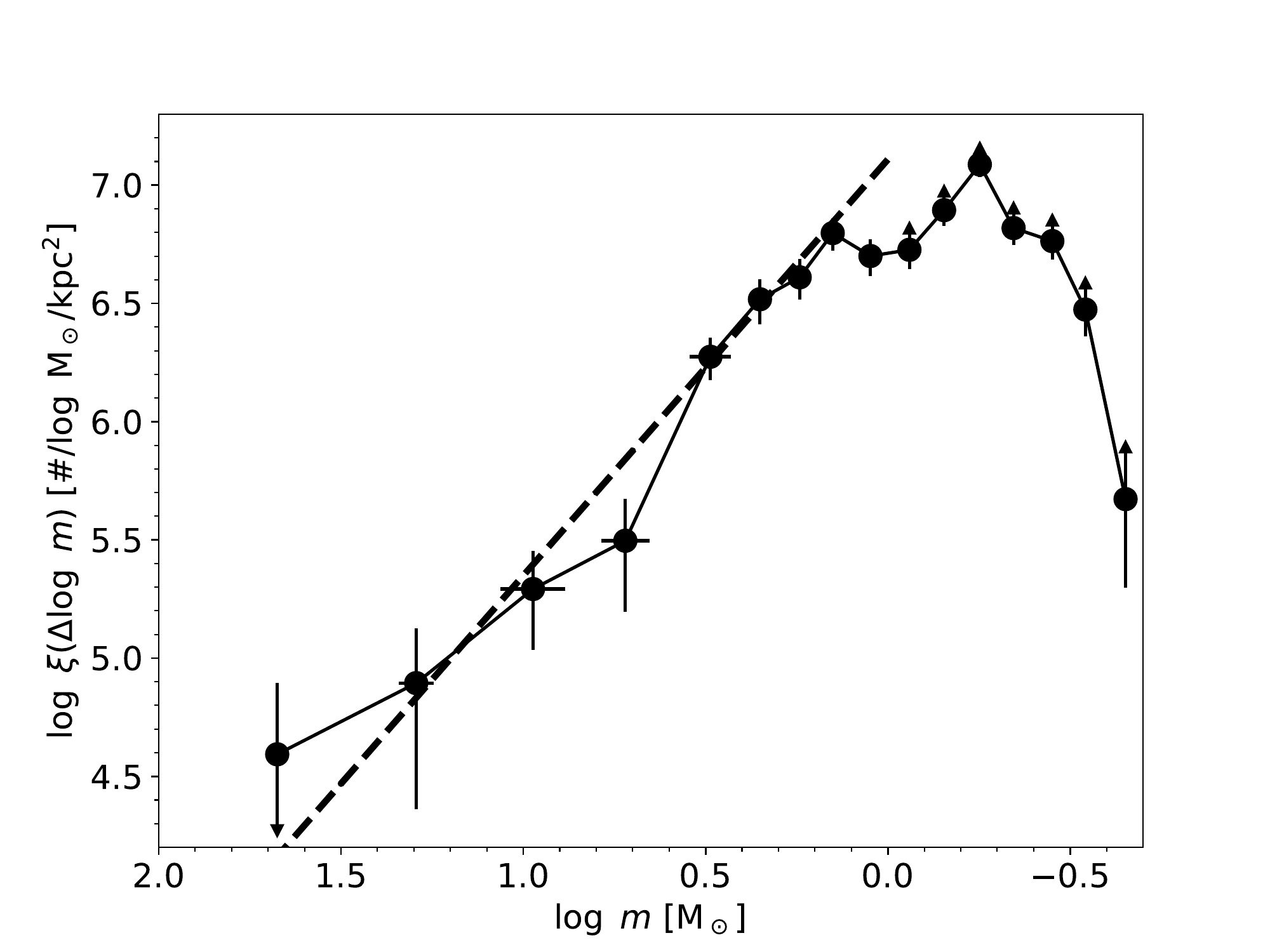}
    \caption{Mass function of IC 1590. The error of $\xi$ is set as $\sqrt{N} \over {\Delta \log m \cdot S}$. The bin sizes are $\Delta \log m = 0.1$ for $\log m < 0.4$, $\Delta \log m = 0.2$ for $0.4 < \log m < 0.8$, and $\Delta \log m = 0.4$ for $\log m > 0.8$. The dashed line shows the best fit, using the maximum likelihood method, to the mass range $m > 1 \rm{M_\odot}$.}
    \label{fig:IMF}
\end{figure}

We derived the IMF in the range from $0.2 ~\mathrm{M_\odot}$ to $60~\rm{M_\odot}$.
Different bin sizes were used to include at least one star in a given mass bin;
$\Delta \log m = 0.1$ for $-0.7 < \log m < 0.4$, $\Delta \log m = 0.2$ for $0.4 < \log m < 0.8$, and $\Delta \log m = 0.4$ for $\log m > 0.8$.
We did not consider the unresolved binary systems.
Figure~\ref{fig:IMF} shows the IMF of IC 1590.
Before calculating the slope of the IMF, we need to limit the stars within the completeness of our photometry and member selection.
Our photometry has a completeness limit of 17.5 mag in the $I$ band, which corresponds to about $1~\rm{M_\odot}$.
Since our member selection depends on X-ray data, the completeness mass of the X-ray dataset should be also considered.
\citet{townsley2019} detected X-ray sources complete down to $0.4 ~\rm{M_\odot}$, and therefore our member selection is likely complete for stars with masses larger than $1 ~\rm{M_\odot}$.

We calculated the slope of the IMF for stars with masses larger than $1~\mathrm{M_\odot}$.
Using the method of maximum likelihood, the slope of the IMF, $\Gamma$ = $-1.49 \pm 0.14$, was determined.
It is slightly steeper than the slope of $-1.35$ derived by \citet{salpeter1955} and \citet{kroupa2001} for the Solar neighborhood.
This implies that low-mass star formation in this SFR is more dominant than that in the Solar neighborhood.
On the other hand, \citet{guetter1997} and \citet{sharma2012} obtained shallow IMFs.
They identified a low number of low-mass members and computed the slope of the IMF for stars with masses larger than $2~\rm{M_\odot}$.
Therefore, the discrepancy between their results and ours presumably originates from the completeness of the low-mass member selection.
%\textcolor{red}{One peculiarity of the IMF is the relatively shallower %slope for $\log m > 0.5$ and an abrupt jump at $\log m \sim 0.5$ which is also seen in the IMF of \citet{sharma2012}.
%\citet{guetter1997} and \citet{sharma2012} might have obtained shallow slopes because their memberships were limited to massive stars.}

\section{Discussion\label{sec:discussion}}
\subsection{The Effect of binarity in IMF}
The effect of binaries has not been considered in the derivation of the stellar IMF \citep{kroupa2001}. However, it is a common phenomenon that most stars form or end up in binary or multiple systems during the star formation process.
A number of previous studies have investigated stellar multiplicity and mass ratio \citep{duquennoy1991, fischer1992, delfosse2004, raghavan2010, sana2012, peter2012}. \citet{duchene2013} reviewed the overall properties of stellar multiplicity. Here, we examined the effects of binarity on the IMF using a Monte-Carlo 
technique.

We constructed a synthetic cluster containing 10000 stars in the mass range of $0.1~\rm{M_\odot}$ to $100 ~\rm{M_\odot}$ using a Monte-Carlo method.
The Kroupa IMF was adopted as the underlying IMF.
The binary fraction was determined as a function of primary mass.
The mass ratio distribution in a given mass bin is expressed as a power law relation $f(q) \propto q^\gamma$ where $q$ is the mass ratio.
The functions for binary fraction and mass ratio were adopted from those of \citet{duchene2013}.
The typical binary fraction of low-mass stars and $\gamma$ are about 20\% and 4.2, respectively.
More than 80\% of massive stars occur in binary systems \citep{sana2012}, and their mass ratio distribution appears to be flat.

\begin{figure}
    \centering
    \includegraphics[width=\columnwidth]{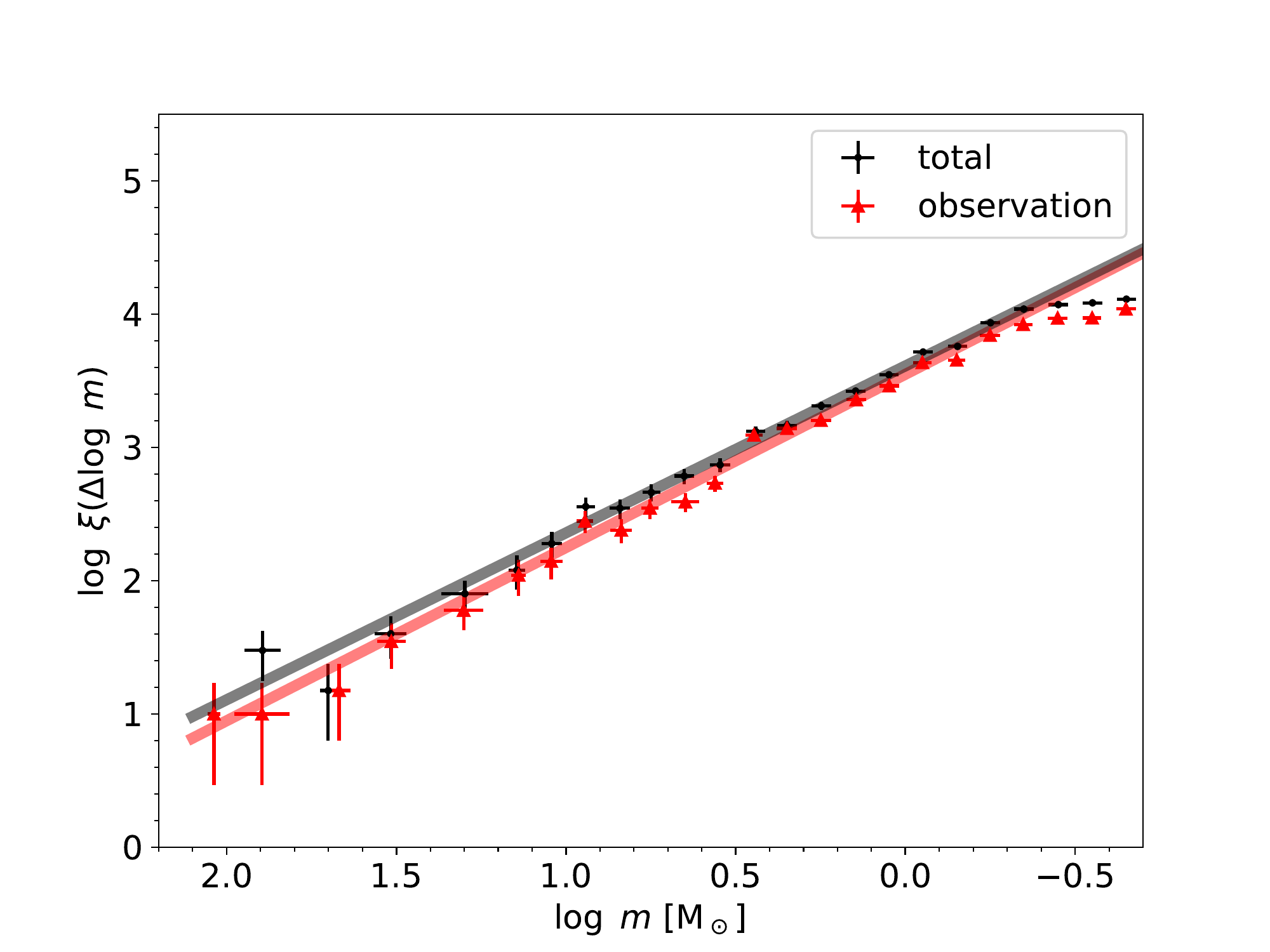}
    \caption{Initial mass functions of the synthetic cluster. Black dots represent the underlying IMF using the resolved binary stars. Red dots represent the IMF for the synthetic cluster when binary systems are considered as single stars. Lines show the slope of the IMF, determined in the mass range $m > 1 ~\rm{M_\odot}$.}
    \label{fig:synthetic_cluster}
\end{figure}

A total of 2730 binary systems and 7270 single stars were generated.
The effective temperature and bolometric magnitude of stars in the synthetic cluster were obtained by interpolating their masses to the evolutionary models for MS and PMS stars \citep{ekstrom2012, siess2000}.
All the binary systems were assumed to be unresolved binaries.
The temperatures of the binary systems may be close to those of the primary stars, while their total luminosities can be obtained from the sum of luminosities of the primary and secondary stars.
In turn, we derived the mass of stars from the HRD using the same method described in Section~\ref{sec:initial_mass_function}.
The IMF of the synthetic cluster was finally derived as shown in Figure~\ref{fig:synthetic_cluster} (red dots).
We also present the underlying IMF taking into account the resolved binary stars (black dots) for comparison.
The slopes of the underlying IMF and simulated one are $-1.25 \pm 0.04$ and $-1.30 \pm 0.04$, respectively.
There is no significant difference between two IMFs.
Therefore, the IMF of IC 1590 may not be significantly affected by the binary population in the cluster.

We can guess some reasons that unresolved binaries do not affect the estimation of the IMF very much.
The primary stars essentially dominate the total luminosities of binary systems.
In the high-mass regime, the size of the mass bins is sufficiently large to cover a number of stars even if stellar masses were misderived due to the contribution of the secondary stars.
The low-mass PMS stars are evolving along the Hayashi tracks, and therefore their masses may not significantly misderived even if their luminosities are increased by the binary effects.

\subsection{Mass Segregation}
Some open clusters exhibit the signature of mass segregation (e.g., the concentration of massive stars; \citealp{sung2013, sung2017, dib2018, dib2019, hetem2019}).
Mass segregation is the result of energy equipartition among cluster members \citep{binney1987}.
This dynamical process takes place on a relaxation time scale which for a typical cluster with about 1000 members is longer 
than ten crossing times \citep{bonnell1998}.
However, because young open clusters are too young to be relaxed, the cause of mass segregation among young open clusters is still debated.
One possibility is that the mass segregation found in several young open clusters could have a primordial origin related to star formation in inhomogeneous molecular clouds \citep{bonnell1998}.
On the other hand, some assert that the origin of mass segregation in young open clusters is due to rapid dynamical evolution in their early stages \citep{mcmillan2007, allison2009, allison2010}. 
Such rapid dynamical evolution occurs when substructures in molecular clouds merge together into a star cluster.
In this situation, mass segregation can dynamically occur on a shorter timescale than the relaxation time.

\begin{figure}
    \centering
    \includegraphics[width=\columnwidth]{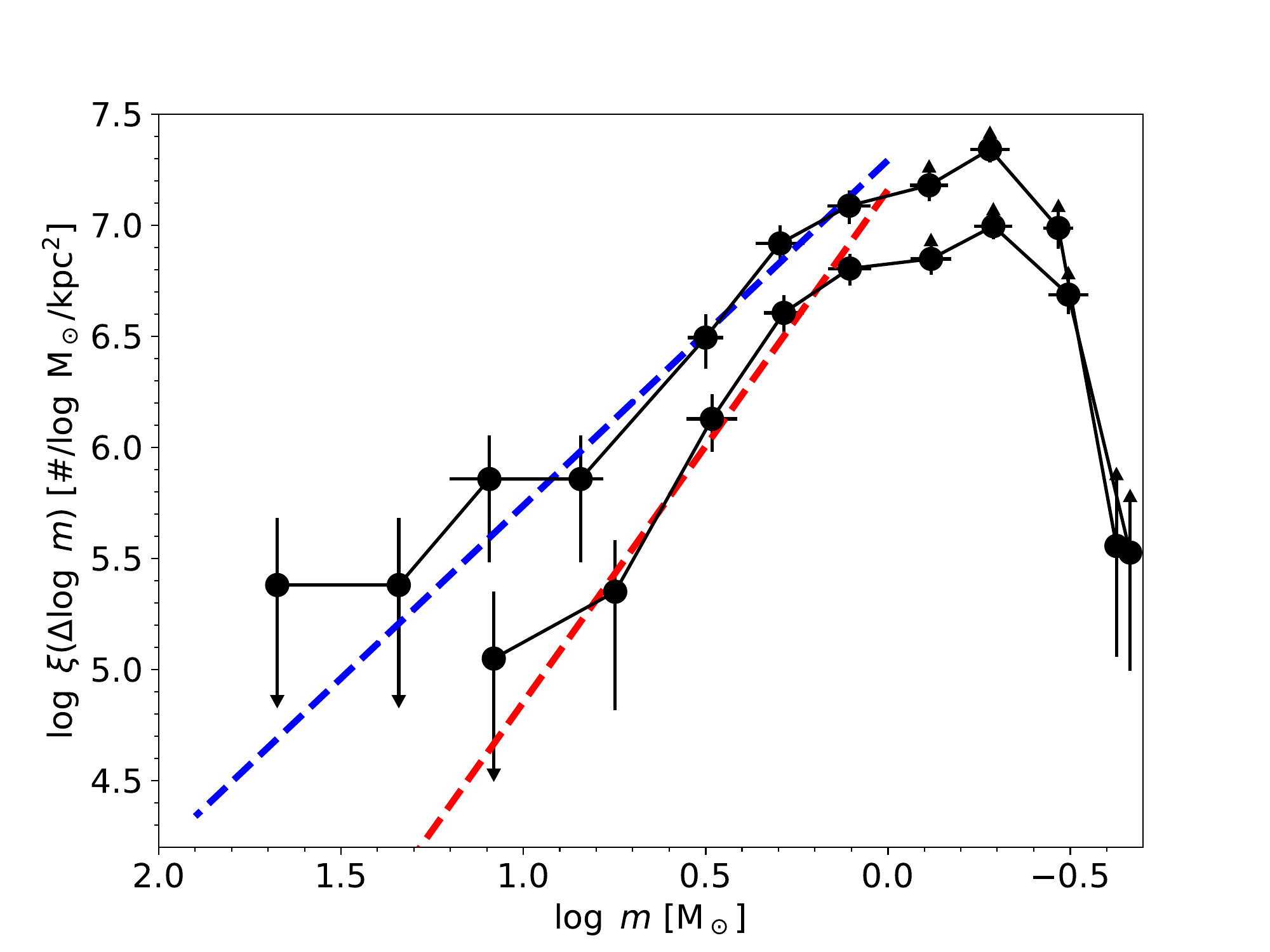}
    \caption{Mass functions for the inner (blue) and outer (red) regions from the half-number radius of $2.^\prime5$. The slopes are determined in the mass range $1 < m/M_\odot < 10$ and $-1.55 \pm 0.27$ for the inner region and $-2.31 \pm 0.21$ for the outer region.}
    \label{fig:cluster_size}
\end{figure}

To investigate the signature of mass segregation in IC 1590, we derived the mass functions (MFs) of stars in the inner and outer regions from the half-number radius of $2.^\prime5$, respectively.
We derived IMFs of stars in the same manner described in Section~\ref{sec:initial_mass_function}.
The number of stars is 206 for the inner region and 202 for the outer region.
Six out of seven massive stars including HD 5005AB reside in the inner region.
Because there is only one star with $m > 10~\rm{M_\odot}$ in the outer region, we determined the slope of the MFs using stars with $1~\rm{M_\odot} < m < 10~\rm{M_\odot}$ for a fair comparison.
The slopes of the MFs are somewhat different; $\Gamma = -1.55 \pm 0.27$ for the inner region and $\Gamma = -2.31 \pm 0.21$ for the outer region, respectively (Figure~\ref{fig:cluster_size}).
This result indicates that mass segregation occurs in IC 1590.

\begin{figure}
    \centering
    \includegraphics[width=0.5\textwidth]{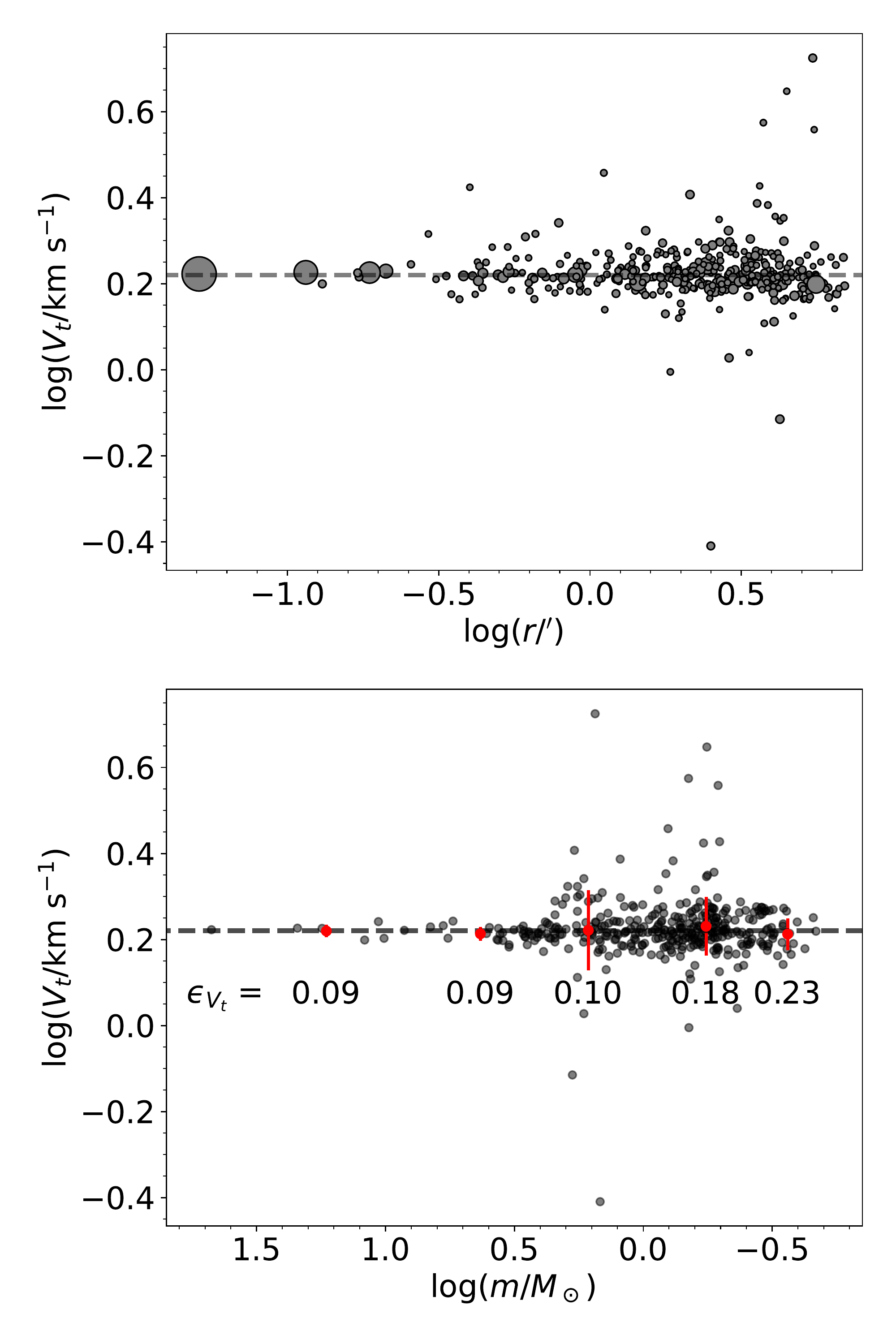}
    \caption{Tangential velocity $V_t$ distribution of cluster members with respect to the radial distance (upper) and stellar mass (lower). In the upper panel, the size of dots is proportional to the mass of individual stars. The dashed lines represent the median $V_t$. In the lower panel, the average and standard deviation of $V_t$ in several mass bins are represented by red dots and error bars, respectively. It is worth noting that the typical uncertainties of tangential velocity $\epsilon_{V_t}$ in a given mass bin are large in low-mass part.}
    \label{fig:tangential_velocity}
\end{figure}

%However, the origin of the mass segregation cannot be judged from this analysis alone.
Were the mass segregation caused by rapid dynamical evolution at an early stage, there would be a signature in the stellar mass and velocity correlation.
Adopting a distance of 2.8 kpc, we computed two-dimensional tangential velocities ($V_t = \sqrt{V^2_{R.A.} + V^2_{Dec.}}$) of members using the proper motions from the {\it Gaia} EDR3 \citep{gaia2020}.
Figure~\ref{fig:tangential_velocity} displays the $V_t$ distribution with respect to the radial distances from the cluster center and stellar masses.
We could not find any mass-dependent correlation.
The dispersion in $V_t$ increases with stellar mass, but the large errors in the proper motions of faint stars may be responsible for the large scatter in $V_t$.
Hence, these results suggest that the mass segregation in IC 1590 more likely has a primordial origin, rather than resulting from the violent dynamical evolution of the cluster.

\subsection{Triggered or Sequential Star Formation}
\begin{figure}
    \centering
    \includegraphics[width=\columnwidth]{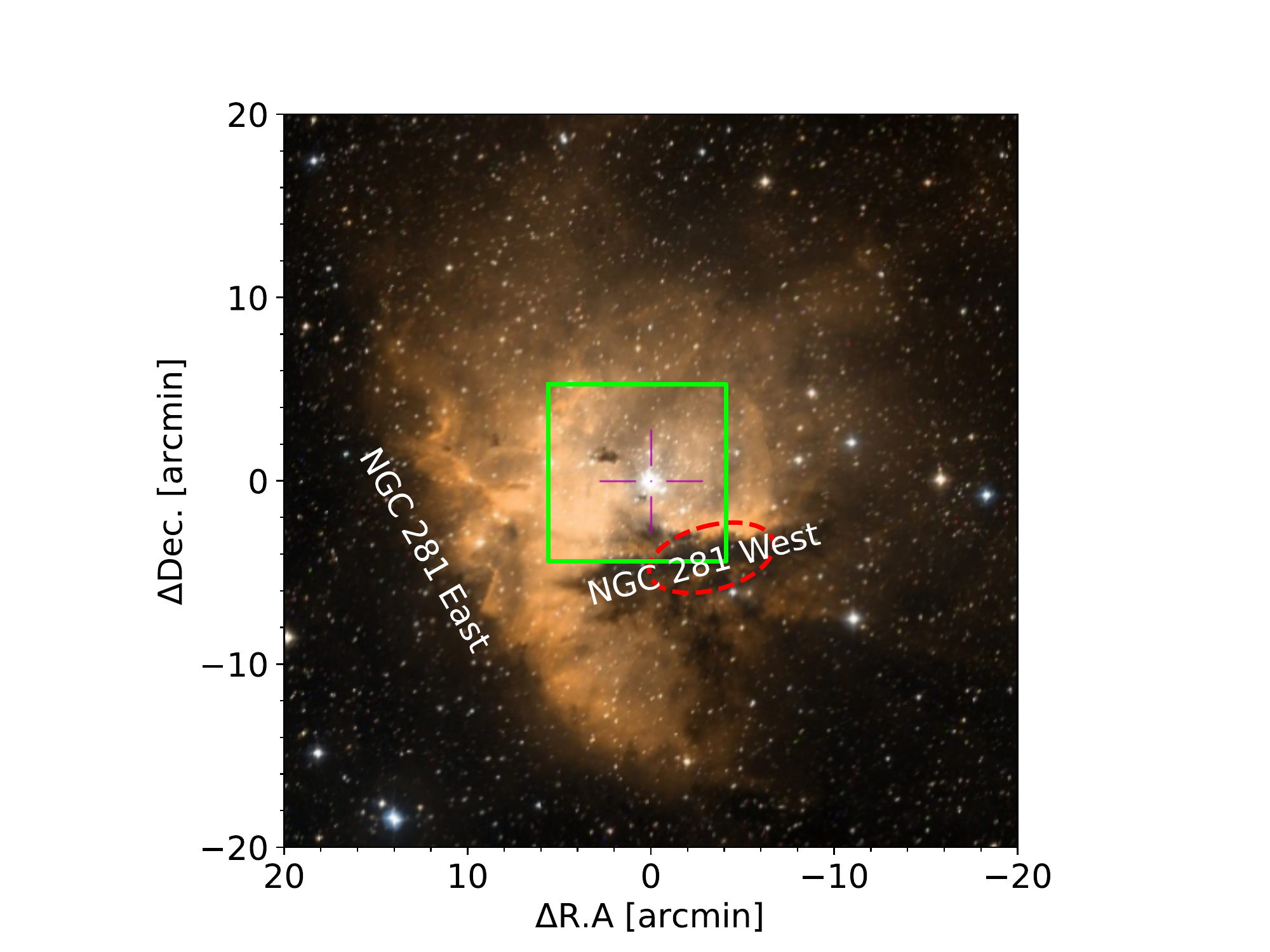}
    \caption{Optical image of IC 1590 and two SFRs taken from the Digital Sky Survey 2. The green rectangular and red ellipse show the FoV of Kuiper 61$^{\prime\prime}$ telescope observation and the northern subcluster of NGC 281 West.}
    \label{fig:full_map}
\end{figure}

Figure~\ref{fig:full_map} displays two SFRs in NGC 281 region.
Based on CO observations and the age sequence of YSOs, \citet{megeath1997} and \citet{sharma2012} suggested that the star formation of NGC 281 West and NGC 281 East were affected by IC 1590.
To investigate the possibility of triggered star formation of NGC 281 West, we estimated the expanding velocity of the HII region NGC 281.
If we assume that the Lyman continuum photons from O-type stars are used only for expanding the HII region, the expansion velocity is given by
\begin{equation}
    v = { N_{LyC} \over {4 \pi r^2 n_H} }
\end{equation}
where $N_{LyC}$ and $n_H$ are the number of the Lyman continuum photons per second and the number density of hydrogen atoms.
\citet{lim2018} derived an expansion velocity for the NGC 1893 HII bubble of $8.6 ~\rm{km~s^{-1}}$.
NGC 1893 contains the same number of O-type stars and has a HII bubble with the similar diameter to NGC 281.
We compared the $N_{LyC}$ from O-type stars and the cluster size of IC 1590 with those of NGC 1893 where $N_{LyC}$ was obtained from \citet{draine2011}.
The expanding velocity of NGC 281 is comparable to that of NGC 1893 when the hydrogen number densities $n_H$ of the two regions are the same.

The projected distance between the O-type Trapezium system of IC 1590 and the northern subcluster of NGC 281 West is about $4.1~\rm{pc}$ at $d = 2.8 ~\rm{kpc}$ (Figure~\ref{fig:full_map}).
The travel time of the ionization front to the northern subcluster of NGC 281 West is 0.5 Myr at the expanding velocity of $8.6~\rm{km~s^{-1}}$.
If IC 1590 has indeed triggered the star formation of NGC 281 West, the age of NGC 281 West would be 0.3 Myr.
Then most of YSOs in NGC 281 West would be at Class 0/I stage if we consider that the lifetime of Class 0/I YSOs is about $0.55 ~\rm{Myr}$ \citep{evans2009}.
However, \citet{sharma2012} detect 13 Class II YSOs in the northern subcluster of NGC 281 West.
Hence, it is not clear whether the HII region driven by massive stars in IC 1590 has triggered the formation of the northern subcluster in NGC 281 West.

\citet{megeath2002, megeath2003} also suggest that the NGC 281 complex was formed in a fragmenting superbubble.
Based on VLBA parallax and proper motion data of the IRAS sources, \citet{sato2008} suggested that the 3-D structure of the superbubble is a slightly elongated shape along the Galactic disk.
They proposed that the first-generation of OB stars in IC 1590, which were formed by the superbubble shock, ionized the surrounding materials and could have influenced the next-generation of star formation in the neighboring clouds NGC 281 West.
If IC 1590 did not trigger the star formation of NGC 281 West, then the star formation of NGC 281 West might have been sequentially affected by the external stimulus such as the shock of superbubble.

\subsection{Individual Notable Stars\label{sec:individual}}
\subsubsection{No. 409}
No. 409 is an early-type star (see the TCD in Figure~\ref{fig:BV_UB}).
This star has almost the same proper motion and parallax of IC 1590.
Also, No. 409 shows strong H$\alpha$ emission and a NIR excess, which implies the existence of an actively accreting hot disk.
Therefore, No. 409 seems to be a Herbig Be star.
We excluded No. 409 in constructing the extinction map.

\subsubsection{No. 1210 and No. 1254}
There is no definitive evidence for the membership of No. 1254 because its proper motion ($\mu_{\alpha}\cos\delta, \mu_\delta$) = ($0.197 \pm 0.012, -3.221 \pm 0.014$) largely deviates from the cluster's value and the star is not an X-ray source nor an H$\alpha$ emitter.
No. 1210 has  a similar proper motion and parallax as members but X-ray emission or H$\alpha$ emission were not detected.
These two sources follow a normal interstellar extinction law, but have similar $E(B-V)$ with other members of IC 1590 in Figure~\ref{fig:reddening}.
Furthermore, they follow the cluster's sequence in Figure~\ref{fig:BV_UB} so they are probable members of IC 1590.
The normal extinction law for No. 1210 and No. 1254 implies that the outer region of NGC 281 West follows a normal interstellar extinction law.

\subsubsection{HD 5005AB}
The most massive stars HD 5005AB (=No. 380) in IC 1590 is too bright to be de-convolved.
The estimated mass of the system is $47\pm7~\mathrm{M_\odot}$.
\citet{sota2011} reported the spectral types of HD 5005A and HD 5005B as O4V((fc)) and O9.7II-III, respectively.
However, it is not consistent with the binary formation scenario and the stellar evolution models for the less massive star HD 5005B to have already evolved away from the main-sequence while the more massive star HD 5005A is still on the main-sequence.
\citet{sota2011} also reported that the luminosity class of HD 5005B from HeII $\lambda4686$/HeI $\lambda4713$ conflicts with a very weak Si IV line.
Therefore, a luminosity class V may be more appropriate for HD 5005B.

\section{Summary\label{sec:summary}}

In this paper, we presented deep $UBVI$ and H$\alpha$ photometry for the central $9.^\prime7 \times 9.^\prime7$ area of the young open cluster IC 1590.
%\textcolor{red}{(about 8 arcmin (Sharma et al. (2012), at least 7 arcmin (Guetter & Turner (1997))}
We identified 39 H$\alpha$ emission stars and 15 H$\alpha$ emission candidates from H$\alpha$ photometry.
We selected 408 members from optical TCDs, CMDs, H$\alpha$ photometry, X-ray emission, {\it Gaia} EDR3 parallax, and proper motion.
Our photometry is complete down to $1~\rm{M_\odot}$.

The mena reddening was estimated to be $<E(B-V)> = 0.40 \pm 0.06$ (s.d.), and there is differential reddening across the cluster.
The extinction law toward IC 1590 was ionvestigated by using the color excess ratios.
As a result, we confirmed an abnormal extinction for the intracluster medium ($R_{V,cl} = 0.39 \pm 0.34$), while a normal extinction law was found for the foreground medium toward the Perseus arm.
The distance modulus of IC 1590 determined from the reddening-free CMDs is $12.3 \pm 0.2 ~\mathrm{mag}$ corresponding to $d = 2.88 \pm 0.28 ~\mathrm{kpc}$.
This result is in good agreement with the distance derived from the {\it Gaia} EDR3.

The ages and masses of members were obtained from stellar evolution models and PMS evolutionary tracks.
The representative age obtained from the distribution of PMS stars is $0.8 ~\mathrm{Myr}$ and the ages of PMS stars distributes from 0.7 Myr (10\% percentile) to 8.4 Myr (90\% percentile).
Our estimation is smaller than the upper limit of 4.4 Myr estimated by \citet{sharma2012}.
The IMF of IC 1590 was derived, and its slope is $\Gamma = -1.49 \pm 0.14$.
It is steeper than Salpeter/Kroupa IMF (-1.35), implying that low-mass star formation is dominant in this SFR.

To find a signature of mass segregation in the IC 1590 region, we compared the IMF for the inner region ($r < 2.^\prime 5$) with that for the outer region ($r > 2.^\prime 5$).
The slope of the IMF for the outer region appears steeper than that for the inner region.
In addition, the distribution of massive stars with $m > 8~\rm{M_\odot}$ is concentrated in the inner region.
These observations indicate the signature of mass segregation in IC 1590.
But, we could not find any evidence for the energy equipartition among members in the correlation between stellar masses and tangential velocities.

To investigate the possibility of triggered star formation of NGC 281 West, we estimated an expansion velocity of the HII region NGC 281 of $8.6 ~\rm{km~s^{-1}}$.
If the O-type stars of IC 1590 have triggered the star formation of the northern sub-cluster of NGC 281 West, the age of the northern sub-cluster would be 0.3 Myr.
However, since there are a number of Class II YSOs probably older than 0.55 Myr in NGC 281 West, it is not evident that the massive stars of IC 1590 triggered the star formation of NGC 281 West.
Also, to study further star formation history of NGC 281 region, another star forming region NGC 281 East \citep{elmegreen1978, megeath1997} and several molecular clouds nearby NGC 281 \citep{lee2003} should be investigated thoroughly.

\acknowledgments
The authors thank the anonymous referee for many constructive comments and suggestions.
We dedicate this work to our colleague Hwankyung Sung, who passed away during the final stages of this investigation.
This work was supported by the National Research Foundation of Korea (NRF) grant funded by the Korean government (MSIT) (Grant No: NRF-2019R1A2C1009475 and NRF-2019R1C1C1005224). 
This work has made use of data from the European Space Agency (ESA) mission Gaia (http://www.cosmos.esa.int/gaia), processed by the Gaia Data Processing and Analysis Consortium (DPAC, http://www.cosmos.esa.int/web/gaia/dpac/consortium). Funding for the DPAC has been provided by national institutions, in particular the institutions participating in the Gaia Multilateral Agreement. The Digitized Sky Surveys were produced at the Space Telescope Science Institute under U.S. Government grant NAG W-2166. The images of these surveys are based on photographic data obtained using the Oschin Schmidt Telescope on Palomar Mountain and the UK Schmidt Telescope. The plates were processed into the present compressed digital from with the permission of these institutions. IRAF was distributed by the National Optical Astronomy Observatory, which was managed by the Association of Universities for Research in Astronomy (AURA) under a cooperative agreement with the National Science Foundation.


\vspace{5mm}
\facilities{Maidanak:1.5m, SO:Kuiper}


\software{IRAF}



%\appendix


%나중에 제목 다 넣을것

%\bibliography{bib.bib}{}
\bibliographystyle{aasjournal}



\end{document}